\documentclass[twocolumn]{aastex63}

\usepackage{latexsym}
\usepackage{amsmath}
\usepackage{graphicx}
\usepackage{longtable,booktabs}
\usepackage{natbib} 

\begin{document}

\title{Unveiling the progenitors of a population of likely peculiar GRBs}

\correspondingauthor{Pak-Hin Thomas Tam}
\email{tanbxuan@mail.sysu.edu.cn}

\author[0009-0007-0686-3906]{Si-Yuan Zhu}
\affiliation{School of Physics and Astronomy, Sun Yat-Sen University, Zhuhai 519082, People's Republic of China}

\author[0000-0002-1262-7375]{Pak-Hin Thomas Tam}
\affiliation{School of Physics and Astronomy, Sun Yat-Sen University, Zhuhai 519082, People's Republic of China}

\begin{abstract}
Traditionally, gamma-ray bursts (GRBs) are classified as long and short GRBs, with $T_{90} = 2$ s being the threshold duration.
Generally, long-duration GRBs (LGRBs, $T_{90}>2$ s) are associated with the collapse of massive stars, and short-duration (SGRBs, $T_{90}<2$ s) are associated with the compact binary mergers involving at least one neutron star.
However, the existence of a population of so-called ``peculiar GRBs", i.e., LGRBs originating from mergers, or long Type I GRBs, and SGRBs originating from collapsars, or short Type II GRBs, have challenged the traditional paradigm of GRB classification. 
Finding more peculiar GRBs may help to give us more insight into this issue.
In this work, we analyze the properties of machine learning identified long Type I GRBs and short Type II GRBs candidates, long GRBs-I and short GRBs-II (the so-called ``peculiar GRBs").
We find that long GRBs-I almost always exhibit properties similar to Type I, which suggests that the merger may indeed produce GRBs with $T_{90}>2$ s.
Furthermore, according to the probability given by the redshift distribution, short GRBs-II almost exhibit properties similar to Type II.
This suggests that the populations of short Type II GRBs are not scarce and that they are hidden in a large number of samples without redshifts, which is unfavorable to the interpretation that the jet progression leads to a missed main emission.
\end{abstract}


\section{Introduction} \label{sec:introduction}
Gamma-ray bursts (GRBs) are the brightest electromagnetic explosions in the universe.
Theoretically, the merger of compact binaries (Type I GRB) and the collapsar of massive stars (Type II GRBs) will produce GRBs \citep{1986ApJ...308L..43P,1993ApJ...405..273W,2007ApJ...655L..25Z,2009ApJ...703.1696Z}.
According to the collapsar scenario, the duration of GRBs is defined by the envelope fallback timescale, typically 10 s \citep{1999ApJ...524..262M}. 
Furthermore, the timescale of GRBs originating from neutron star--neutron star (NS--NS) and neutron star--black hole (NS--BH) mergers typically is 0.01--0.1 s, which suggests that merged GRBs should typically have short durations \citep{2005A&A...436..273A}.

According to the bimodal distribution of durations, GRBs are directly classified into long GRBs (LGRBs, $T_{90}>2$ s) and short GRBs (SGRBs, $T_{90}<2$ s) \citep{1993ApJ...413L.101K}, where $T_{90}$ is time interval with integrated photon counts raise from 5\% to 95\%.
Previously, observations have revealed that LGRBs are associated with supernovae (SNe) originating from massive collapsars (such as GRB 030329/SN 2003dh) \citep{2003ApJ...591L..17S}, and SGRBs are associated with gravitational waves (GWs) and kilonova (KNe) both originating from compact binary mergers (such as GRB 170817A/GW170817) \citep{2017ApJ...848L..13A,2017PhRvL.119p1101A,2017ApJ...851L..18W}.

GRBs originating from different progenitors exhibit distinct observational properties.
Spectral lag ($\tau$) is defined as the time delay of high-energy photons with respect to low-energy photons \citep{1986ApJ...301..213N,1995A&A...300..746C}.
Generally, LGRBs have significant $\tau$, while SGRBs have negligible or tiny $\tau$ \citep{2006MNRAS.367.1751Y,2015MNRAS.446.1129B}.
Furthermore, \cite{2002AA...390...81A} analyzed 10 LGRBs with known redshift (z) and first found a tight correlation between the rest frame peak energy ($E_{\rm p,z}$) in the $\nu f_\nu$ spectrum and isotropic energy ($E_{\rm iso}$), the $E_{\rm p,z}$--$E_{\rm iso}$ correlation.
LGRBs and SGRBs follow different $E_{\rm p,z}$--$E_{\rm iso}$ correlation, and exhibit significant separation in the $E_{\rm p,z}$--$E_{\rm iso}$ plane \citep{2013MNRAS.430..163Q,2020MNRAS.492.1919M,2023MNRAS.524.1096L,2023ApJ...950...30Z}.

However, recent observations suggest that the dichotomy based on phenomenological does not necessarily correspond to the two progenitors of GRBs.
GRB 211211A is a LGRB with $T_{90}$ $\sim$ 34 s.
Its lightcurve of whole emission (WE) is characterized by an initial short-hard spiky main emission (ME) followed by a long-soft extended emission (EE), similar to some SGRBs with EE associated with KNe \citep{2016NatCo...712898J,2017ApJ...837...50G}.
In addition, Its optical and near-infrared transients that emerged after the burst were similar to KN after GW170817, which suggests that it is a Type I GRB \citep{2022Natur.612..223R,2022Natur.612..228T,2022Natur.612..232Y}.
\cite{2022Natur.612..232Y} calculated $\tau$ of both ME and EE and found their tiny $\tau$ is consistent with typical SGRBs.
In the $E_{\rm p,z}$--$E_{\rm iso}$ plane, ME follows the track of SGRBs, while EE and WE do not \citep{2022Natur.612..232Y}.
Although the WE duration is mostly contributed by EE for GRB 211211A, the ME duration ($\sim$ 13 s) is still too long for merger scenario.
The neutron star--white dwarf (NS--WD) merger was suggested as an explanation that theoretically allows it to produce long duration GRBs \citep{2007MNRAS.374L..34K,2022Natur.612..232Y,2023ApJ...947L..21Z}.
However, the collapsar origin has not been ruled out \citep{2023ApJ...947...55B}.

GRB 200826A is a SGRB with $T_{90}$ $\sim$ 1 s but associated with a SN, suggesting it is a Type II GRB \citep{2021NatAs...5..917A,2021NatAs...5..911Z,2022ApJ...932....1R}.
\cite{2021NatAs...5..911Z} calculated the energy-dependent $\tau$ between the lowest energy (10-20 keV) band and any higher energy band.
It has a maximum $\tau$ of 157 $\pm$ 51 ms, which is consistent with typical LGRBs.
GRB 200826A also follow the $E_{\rm p,z}$--$E_{\rm iso}$ track of LGRBs.
However, its duration is much shorter than the timescale of collapsar scenarios.
Missing the ME phase because of geometric effects caused either by jet precession or companion obstruction models has been proposed to explain GRB 200826A while putting forward a high demand on the rarity of SGRBs originating from collapsars \citep{2022ApJ...931L...2W}.

Besides the above, some SGRBs (such as GRB 090426 and GRB 100816A) are thought to possibly originate from collapsars \citep{2009ApJ...703.1696Z,2011ApJ...739...47F,2011A&A...531L...6N,2011MNRAS.410...27X,2012ApJ...750...88Z}, as well as some LGRBs (such as GRB 060614 and GRB 230307A) are thought to possibly originate from mergers \citep{2006Natur.444.1050D,2006Natur.444.1047F,2006Natur.444.1053G,2006Natur.444.1044G,2024Natur.626..737L,2024Natur.626..742Y}.
They have broken the correspondence between the duration of GRBs and their progenitors.
According to the different observational properties, \cite{2009ApJ...703.1696Z} proposed that GRBs can be classified based on multiband observation.
However, most GRBs cannot be accurately located and follow observation limited by observation.
The contradiction between observation and theory on timescales has not been resolved.
Additional long Type I GRBs and short Type II GRBs are required to support that they are the tail of $T_{90}$ distribution or the result of some special physical process.

Machine learning has been widely used to classify GRBs based solely on lightcurves or physical parameters of the prompt emission ($\gamma$-ray) \citep{2020ApJ...896L..20J,2023MNRAS.525.5204B,2023ApJ...949L..22D,2023ApJ...945...67S,2024MNRAS.527.4272C,2024MNRAS.532.1434Z}.
The unsupervised dimensionality reduction algorithm, t-distributed stochastic neighbor embedding (t-SNE) \citep{2008JMLR.9.2579M,2014JMLR.15.3221M} and Uniform Manifold Approximation and Projection (UMAP) \citep{2018arXiv180203426M}, can reduce adjacent datapoints in high-dimensional space to adjacent points in a two-dimensional space without prior for partial burst classification.
\cite{2024MNRAS.532.1434Z} applied the t-SNE and UMAP to the Fermi Catalog and found two clusters with a clear separation in the t-SNE and UMAP maps, GRBs-I (may correspond to Type I GRBs) and GRBs-II (may correspond to Type II GRBs).
Interestingly, the distribution of these two types overlaps in $T_{90}$. 
With a boundary of 2 s, there are some long GRBs-I and short GRBs-II that may correspond to long Type I GRBs and short Type II GRBs, respectively, which are peculiar GRBs.
In the peculiar GRBs, GRB 200826A is classified as GRBs-II along with the typical Type II GRBs, which supports the machine learning classification method that short GRBs-II originate from collapsars and long GRBs-I originate from mergers.
Unfortunately, the progenitor of remaining long GRBs-I and short GRBs-II cannot be identified due to the lack of associated SNe/KNe observations.

In this work, we aim to determine the progenitors of long GRBs-I and short GRBs-II by studying the variation of their properties with redshift, since most of them have no observed redshift.
The structure of our article is organized as follows.
In Section \ref{sec:data}, we describe the sample selection criteria and data analysis.
The properties of the long GRBs-I and short GRBs-II are shown in Section \ref{sec:properties}.
The discussions are shown in Section \ref{sec:discussions}.
The conclusions are shown in Section \ref{sec:conclusions}.

\section{Sample Selection and Data Analysis} \label{sec:data}

In order to investigate the detailed properties of GRBs, accurate spectral parameters are crucial.
The Fermi Gamma-ray Burst Monitor (GBM) detector with the broad energy band (8 keV-40 MeV) has enriched the sample of GRBs with reliable $E_{\rm p}$ \citep{2009ApJ...702..791M}.
The Fermi GRBs are taken from the Fermi Catalog\footnote{https://heasarc.gsfc.nasa.gov/W3Browse/fermi/fermigbrst.html} until the end of April 2023 \citep{2020ApJ...893...46V}.
We mainly consider GRBs with a well-measured spectrum.
The four spectral models including the Band model \citep{1993ApJ...413..281B}, the power law (PL) model, the cutoff power law (CPL) model, and the smoothly broken power law (SBPL) model are used to fit the spectra of GRBs.
The values of the spectral parameter and fluence ($S_{\gamma}$) are mainly taken from the best spectral fitting model, which is obtained directly from the Fermi Catalog.
If the best fitting model of one GRB is the PL model, we take the CPL model to get the spectrum parameter and $S_{\gamma}$.
To ensure the accuracy of $E_{\rm p}$, we exclude the GRBs in which the error of $E_{\rm p}$ is larger than 40\%.
To eliminate the effect of the time resolution on the peak flux ($F_{\rm p}$), we uniformly select the $F_{\rm p}$ with the timescale of 64 ms for all GRBs.
A total of 2057 GRBs are selected.
In order to better identify the progenitors of peculiar GRBs, we add 32 GRBs with SN signals, 7 GRBs with KNe signals (the MEs of GRB 060614, GRB 211211A, and GRB 211227A are recorded as a separate sample), and one GRB may originating from magnetar giant
flare (MGF), a total of 40 GRBs.
Finally, we compiled a sample containing 2097 GRBs, some observation data of GRBs with redshift are listed in Table \ref{table:rest}.

\cite{2024MNRAS.532.1434Z} discovered 42 long GRBs-I and 14 short GRBs-II from the Fermi Catalog through UMAP and discovered 47 long GRBs-I and 14 short GRBs-II from the Fermi Catalog through t-SNE.
We compiled these GRBs as the peculiar sample, a total of 61 GRBs, and divided them into four groups, the details of which are listed in Table \ref{table:paticular}.
Group A contains 41 LGRBs classified as long GRBs-I by both t-SNE and UMAP.
Group B contains 13 SGRBs classified as short GRBs-II by both t-SNE and UMAP.
GRB 200826A was not included in Group B for further analysis since its progenitors were known.
Group C contains 6 GRBs classified as long GRBs-I by t-SNE but they are classified as GRBs-II by UMAP.
Group D contains 1 SGRB classified as short GRBs-II by t-SNE but they are classified as GRBs-I by UMAP.

The hardness ratio (HR, defined as the ratio of the observed counts of high-energy photons to the low-energy photons) of Fermi GRBs that are calculated by the ratio of the observed counts (background subtraction) in the energy range of 50--300 and 10--50 keV from \cite{2020ApJ...893...46V}, except for one of the negative values.
We downloaded the GBM time-tagged event (TTE) data of long GRBs-I and short GRBs-II from Fermi Archive FTP websites\footnote{https://heasarc.gsfc.nasa.gov/FTP/fermi/data/gbm}.
The TTE data from all triggered NaI detectors were used in our analysis.
The lightcurves were extracted with the standard Fermi tool for python source \textit{GBM Data Tools}.

In order to explore the physical properties of GRB in the rest frame, we compiled a GRB sample with redshift and listed in Table \ref{table:rest}, a total of 153 GRBs.
Then we estimated the rest frame parameters of the GRB sample with redshift.
In this paper, both $E_{\rm iso}$ and $L_{\rm iso}$ are corrected to the energy band of 1--$10^{4}$ keV in the rest frame.
The $E_{\rm iso}$ is calculated by 
\begin{equation}\label{Eiso}
	E_{\rm iso} = \frac{4\pi D_{\rm L}^2 S_{\gamma}k}{(1+z)},
\end{equation}
where $D_{\rm L}$ is the luminosity distance, $S_{\gamma}$ is the fluence, and $k$ is the $k$--correction factor.
In this paper, we assume a flat universe ($\Omega_{\rm k} = 0$) with the cosmological parameters $H_{0} = 71$ km s$^{-1}$ Mpc$^{-1}$, $\Omega_{\rm M} = 0.27$, and $\Omega_{\Lambda} = 0.73$.
The correction factor $k$ is defined as
\begin{equation}\label{k}
	k = \frac{{\int_{1/(1 + z)}^{{10^4}/(1 + z)} {EN(E)dE} }}{{\int_{{e_{\min }}}^{{e_{\max }}} {EN(E)dE} }},
\end{equation}
where $e_{\rm min}$ and $e_{\rm max}$ are the observational energy band of fluence, $N(E)$ denotes the photon spectrum of GRB \citep{2007ApJ...660...16S}.
The best fitting spectral model for $N(E)$ is selected.
The isotropic luminosity $L_{\rm iso}$ is calculated by
\begin{equation}\label{Liso}
	L_{\rm iso} = 4\pi D_{\rm L}^2 F_{\rm p}k,
\end{equation}
where $F_{\rm p}$ is the peak flux, in units of erg cm$^{-2}$ s$^{-1}$.
When GRBs with photon peak flux $P_{\rm p}$ (in units of photon cm$^{-2}$ s$^{-1}$),  $F_{\rm p}$ is calculated by:
\begin{equation}\label{Fp}
	F_{\rm p} = P_{\rm p} \frac{{\int_{{e_{\min }}}^{{e_{\max }}} {EN(E)dE} }}{{\int_{{e_{\min }}}^{{e_{\max }}} {N(E)dE} }}.
\end{equation}
The rest frame peak energy $E_{\rm p,z}$ and duration $T_{\rm 90,z}$ are calculated as $E_{\rm p,z} = E_{\rm p}(1 + z)$ and $T_{\rm 90,z} = T_{90}/(1 + z)$, respectively.
The symbolic notation $Q_{\rm n} = Q/10^{\rm n}$ is adopted.

\section{Properties of the Long GRBs-I and Short GRBs-II} \label{sec:properties}

\subsection{Spectral Hardness} \label{subsec:hardness}
Spectral hardness can usually be represented by $HR$, the low-energy spectral index ($\alpha$), and $E_{\rm p}$.
Generally, SGRBs are harder than LGRBs.
SGRBs and LGRBs are clustered in distinct regions in the $HR$--$T_{90}$, the $\alpha$--$T_{90}$, and the $E_{\rm p}$--$T_{90}$ planes \citep{2012ApJ...750...88Z}.

\begin{figure*}
	\centering
	\includegraphics[angle=0,scale=0.50]{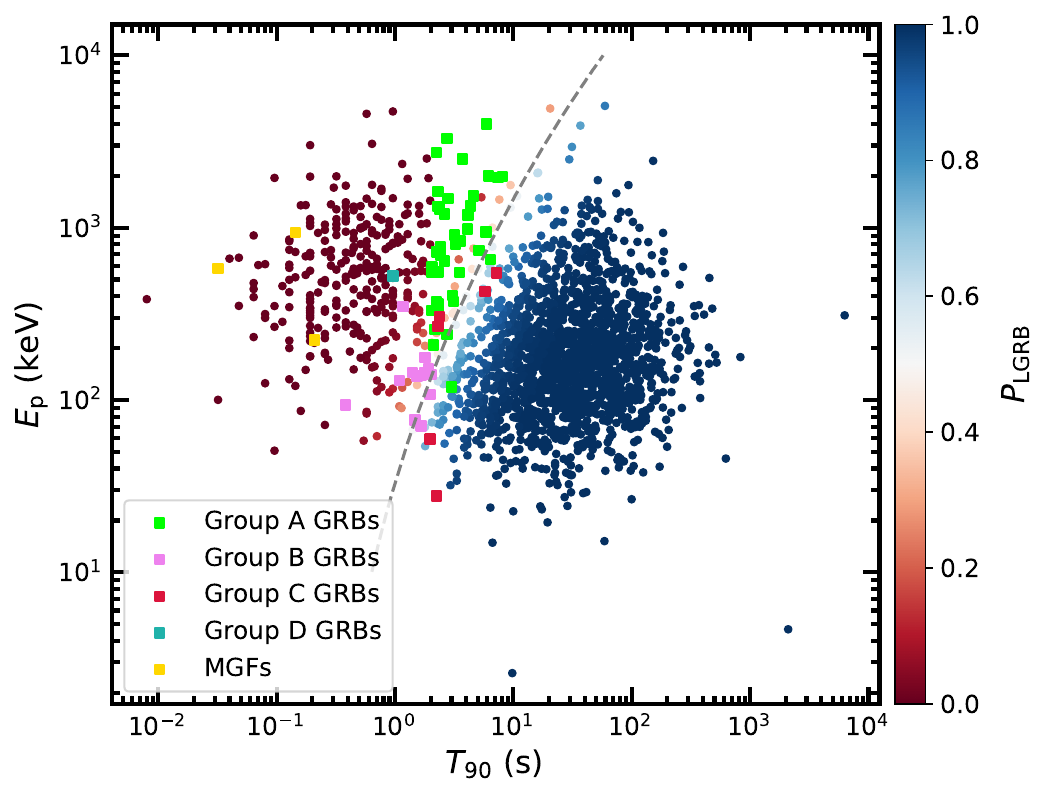}
	\includegraphics[angle=0,scale=0.50]{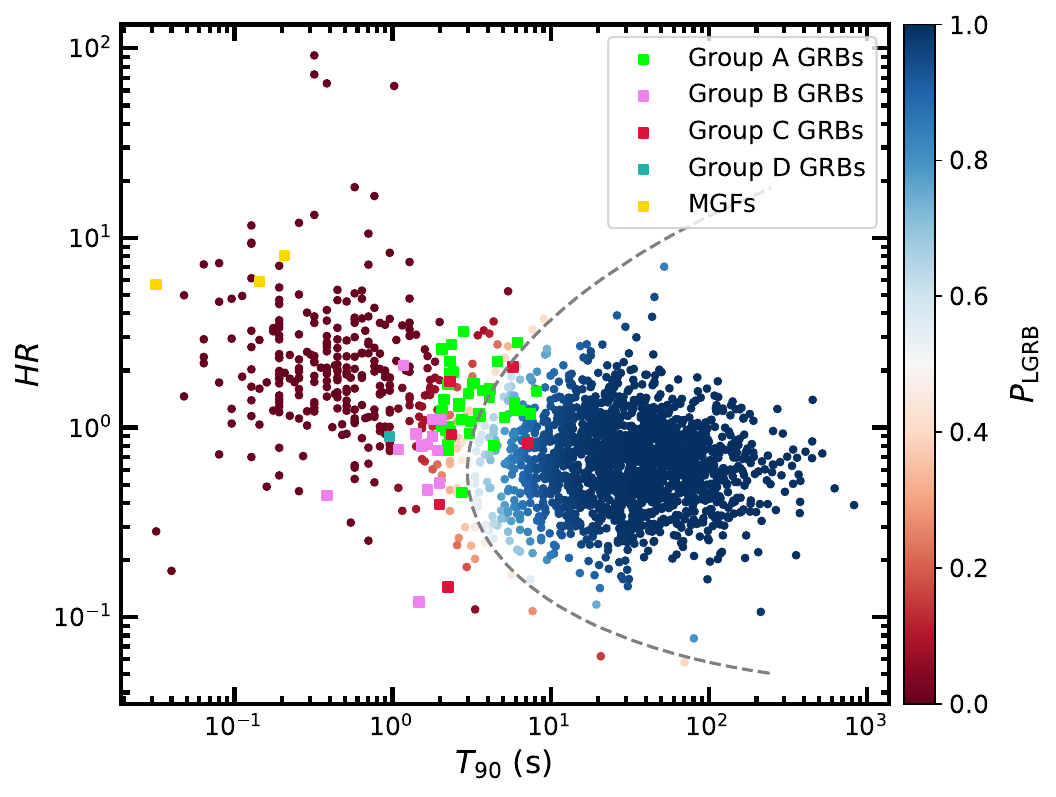}
	\caption{The $E_{\rm p}$--$T_{90}$ plane (left) and the $HR$--$T_{90}$ plane (right). The colour of the data points from red to blue represents the probability of classification as LGRBs from 0 to 100\%. The dashed line is the cutoff where the probability of LGRBs is 50\%. The lime, violet, crimson, and lightseagreen squares represent the Group A, B, C, D GRBs, and MGFs, respectively.}
	\label{figure:hardness}
\end{figure*}

In both the $E_{\rm p}$--$T_{90}$ and the $HR$--$T_{90}$ planes, GRBs of our sample are clustered in distinct regions, as shown in Figure \ref{figure:hardness}.
We use the Gaussian mixture model (GMM) to divide the GRBs into LGRBs and SGRBs and given the probability that each GRB is classified as a LGRB \citep{2014PASJ...66...42T,2016MNRAS.462.3243Z}.
The probabilities of the peculiar GRB samples are listed in Table \ref{table:paticular}.

For Group A, only GRB 080828189 and GRB 081006604 are classified as LGRBs in the $E_{\rm p}$--$T_{90}$ plane with 80.97\% and 62.31\% probability, respectively, and 9 GRBs (including GRB 081006604) are classified as LGRBs in the $HR$--$T_{90}$ plane with less than 90\% probability.
Only four GRBs, GRB 081107321, GRB 100816026 (or, GRB 100816A), GRB 171126235, and GRB 180511437 in Group B are classified as LGRBs in the $E_{\rm p}$--$T_{90}$ plane with less than 65\% probability, and all are classified as SGRBs in the $HR$--$T_{90}$ plane.
GRB 090320045 and GRB 140912664 in Group C are classified as SGRBs in the $E_{\rm p}$--$T_{90}$ plane, and except for GRB 101002279 and GRB 120504945, they are all classified as SGRBs in the $HR$--$T_{90}$ plane.
Group D is classified as SGRBs by different planes.

In addition, some studies have found that MGFs, if occurring in nearby galaxies, would appear as cosmic short-hard GRBs \citep{2020ApJ...899..106Y,2020ApJ...903L..32Z,2021Natur.589..207R,2021Natur.589..211S,2024Natur.629...58M,2024A&A...687A.173T,2024ApJ...963L..10Y}.
The lightcurves and spectrum of MGFs are not significantly different from classical SGRBs.
Compelling evidence for period-modulated tail induced by the magnetar rotation cannot be observed by current wide-field gamma-ray monitors \citep{2021Natur.589..211S}.
Thus, the main evidence for distinguishing MGFs from SGRBs is the spatial coincidence with nearby galaxies and the consistency of the energetics with known MGFs.
When they cannot be localized, the rapid rise in the lightcurve, duration on the order of milliseconds, and very short time variability are possible clues, but not unique \citep{2024A&A...687A.173T}.
We will consider the possibility that MGFs serve as progenitors of peculiar GRBs.
In our sample, three GRBs are considered as MGF candidates, GRB 180128A, GRB 200415A, and GRB 231115A.
They are classified as SGRBs in both the $E_{\rm p}$--$T_{90}$ and the $HR$--$T_{90}$ planes and do not show special features.

\subsection{The $E_{\rm p,z}$--$E_{\rm iso}$ plane} \label{subsec:energy}
Since the $E_{\rm p,z}$--$E_{\rm iso}$ correlation was discovered by \cite{2002AA...390...81A}, classification of GRBs based on the $E_{\rm p,z}$--$E_{\rm iso}$ plane have been widely studied.

\subsubsection{The $E_{\rm p,z}$--$E_{\rm iso}$ correlation} \label{subsubsec:amati}

The $E_{\rm p,z}$--$E_{\rm iso}$ correlation have been widely discussed as a GRB classifier \citep{2012ApJ...750...88Z,2013MNRAS.430..163Q,2020ApJ...903L..32Z,2021NatAs...5..877A,2021NatAs...5..911Z,2022Natur.612..232Y,2022ApJ...936L..10Z,2023ApJ...950...30Z}.
Recent studies have confirmed that the Type I GRBs and the Type II GRBs follow different $E_{\rm p,z}$--$E_{\rm iso}$ correlations, and exhibit significant separation in the $E_{\rm p,z}$--$E_{\rm iso}$ plane, whereas some low-luminosity LGRBs (LL-LGRBs) are inconsistent with the $E_{\rm p,z}$--$E_{\rm iso}$ correlation of Type II GRBs \citep{2006MNRAS.372..233A,2018PASP..130e4202Z,2020MNRAS.492.1919M,2023MNRAS.524.1096L,2023ApJ...950...30Z}.

\begin{figure*}
	\centering
	\includegraphics[angle=0,scale=0.8]{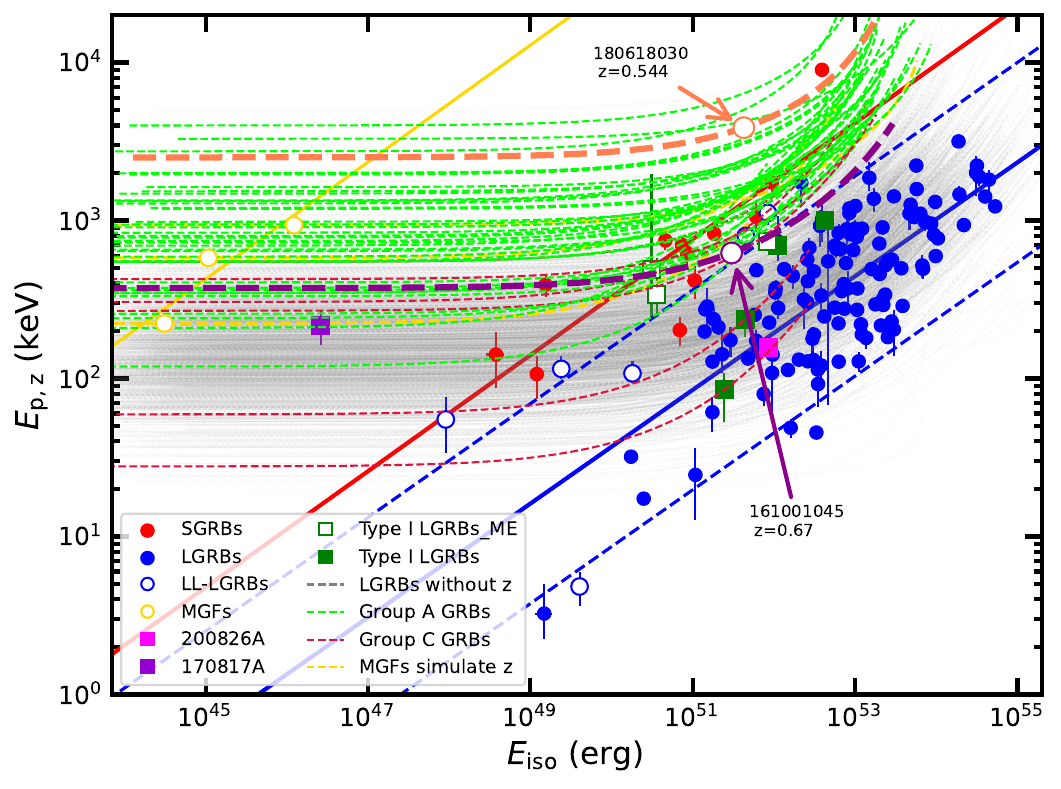}
	\includegraphics[angle=0,scale=0.8]{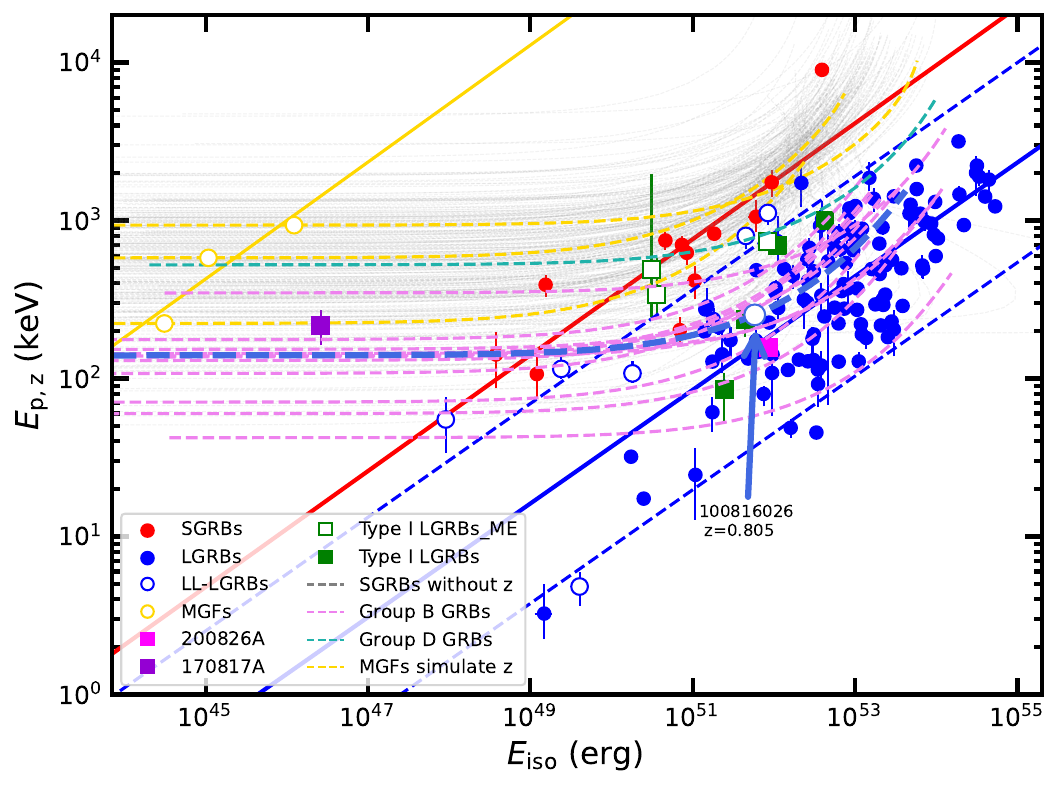}
	\caption{The $E_{\rm p,z}$--$E_{\rm iso}$ plane including the trajectories of Group A, C GRBs and LGRBs without z (top), and Group B, D GRBs and SGRBs without z (bottom). The red and blue solid lines are the best-fit lines for SGRBs and LGRBs fitted by the Markov Chain Monte Carlo (MCMC) method, respectively, and both are $E_{\rm p,z} \propto E_{\rm iso}^{0.37}$. The gold solid line is the translation of $E_{\rm p,z} \propto E_{\rm iso}^{0.37}$. The blue dashed lines are the 2$\sigma$ confidence regions. The lime, violet, crimson, lightseagreen, gold, royalblue, darkmagenta, coral, and gray dashed lines are the trajectories that changes with redshift (gradually increasing from the left to the right) of Group A, B, C, D GRBs, MGFs, GRB 100816026, GRB 161001045, GRB 180618030, and LGRBs (SGRBs), respectively. The gold, royalblue, darkmagenta, coral, and gray rings are MGFs, GRB 100816026, GRB 161001045, and GRB 180618030 at their corresponding redshifts. The red points are SGRBs. The blue points and rings are LGRBs and low-luminosity LGRBs, respectively. The green squares and rings are Type I LGRBs and their MEs.}
	\label{figure:ep-eiso}
\end{figure*}

Almost all of the peculiar GRBs in our sample are without redshift, so we simulate $E_{\rm p,z}$ and $E_{\rm iso}$ for peculiar GRBs at redshifts from 0.0001 to 10, the redshift evolution trajectories are shown in Figure \ref{figure:ep-eiso}.
We can see that Group A GRBs and Group B GRBs are significantly different from the trajectories of LGRBs and SGRBs in the $E_{\rm p,z}$--$E_{\rm iso}$ plane.
Apart from GRB 080828189 and GRB 161001045 (or, GRB 161001A), whose trajectories slightly cross the 2$\sigma$ prediction interval boundary of the $E_{\rm p,z}$--$E_{\rm iso}$ correlation for LGRBs, all of the remaining Group A GRBs are above this boundary, which is clearly different from other LGRBs.
However, the redshifts of GRB 161001045 and GRB 180618030 (or, GRB 180618A) are 0.67 and 0.544, respectively. 
Their confirmed $E_{\rm p,z}$ and $E_{\rm iso}$ indicate that they are both outliers above the 2$\sigma$ prediction interval of LGRBs \citep{2022ApJ...940...56F,2022ApJ...939..106J}.
Group B GRBs enter the 2$\sigma$ prediction interval of LGRBs as the redshift becomes larger, and are within the 2$\sigma$ prediction interval over a wide range.
Among them, GRB 100816026 has $z=0.805$ and follows the track of the $E_{\rm p,z}$--$E_{\rm iso}$ correlation of LGRBs.
Group C GRBs are all above the 2$\sigma$ prediction interval except for GRB 081130212 and GRB 131128629, which are within the 2$\sigma$ prediction interval in the large redshift range.
Group D GRBs are within the 2$\sigma$ prediction interval in the few redshift range.

The MGFs are located in the upper left corner of the $E_{\rm p,z}$--$E_{\rm iso}$ plane, and their tracks with redshift in the $E_{\rm p,z}$--$E_{\rm iso}$ plane are not significantly different from the SGRBs.
This indicates that if the distance or redshift of the SGRBs cannot be determined, MGFs can hardly be recognized.

\begin{figure*}
	\centering
	\includegraphics[angle=0,scale=0.80]{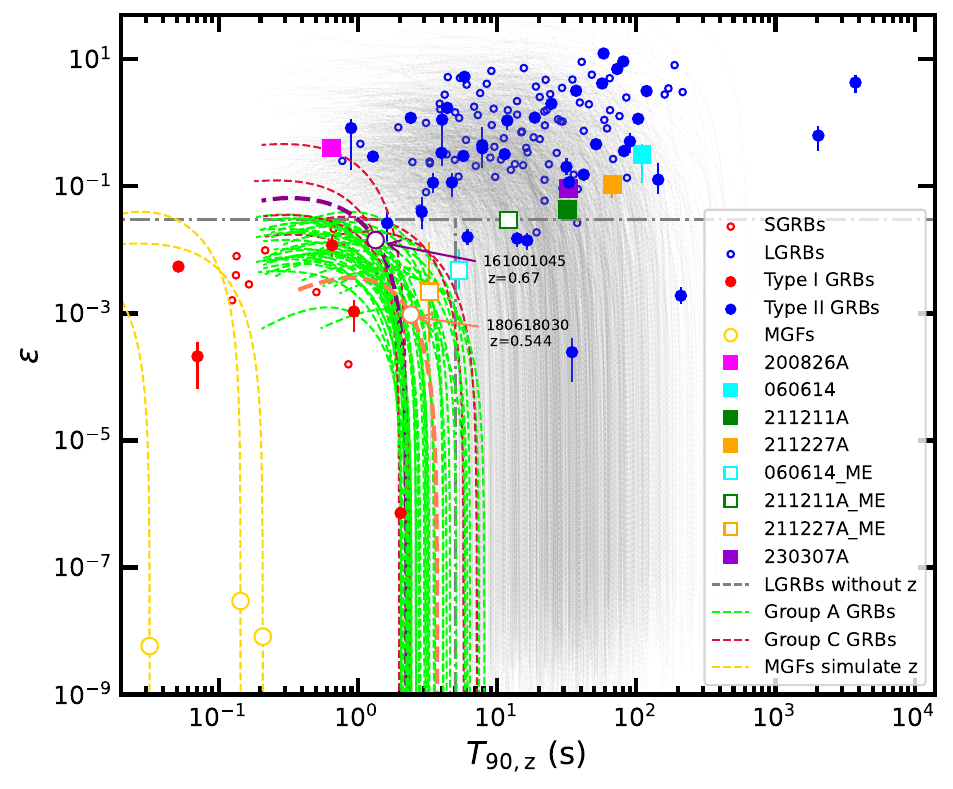}
	\includegraphics[angle=0,scale=0.80]{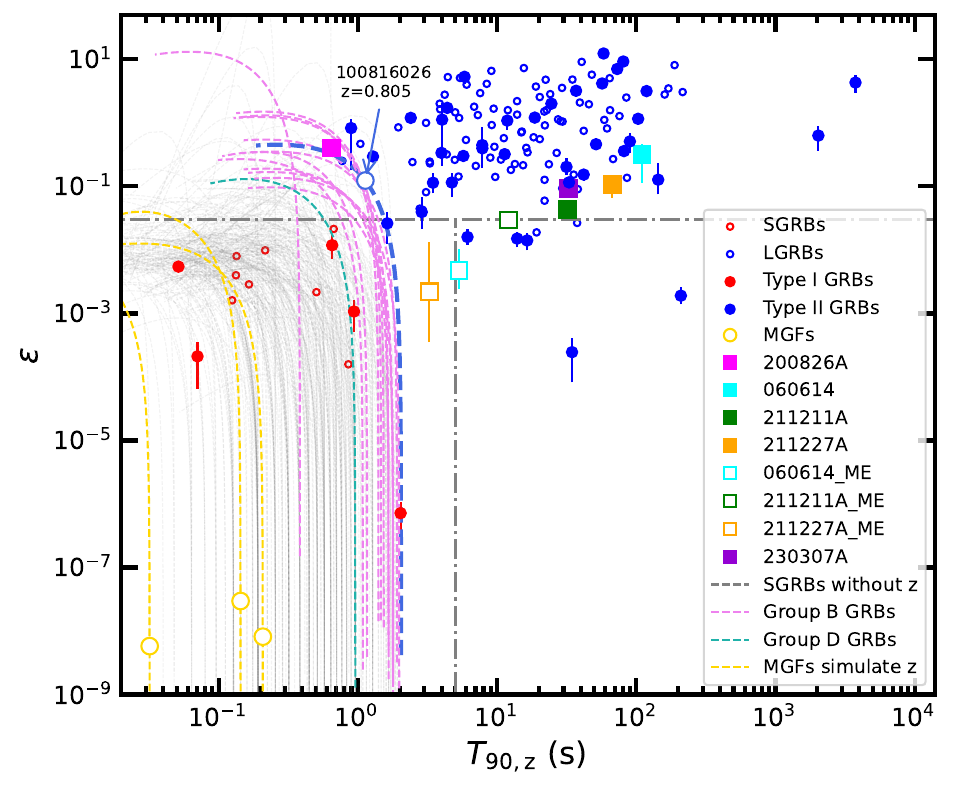}
	\caption{The $\varepsilon$--$T_{\rm 90,z}$ plane. The gray dotted lines are $\varepsilon=0.03$ and $T_{\rm 90,z}=5$ s. The red rings and points are SGRBs and Type I GRBs. The blue rings and points are LGRBs and Type II GRBs. The redshift increases gradually from the bottom right to the top left.}
	\label{figure:epsilon}
\end{figure*}

\begin{figure*}
	\centering
	\includegraphics[angle=0,scale=0.50]{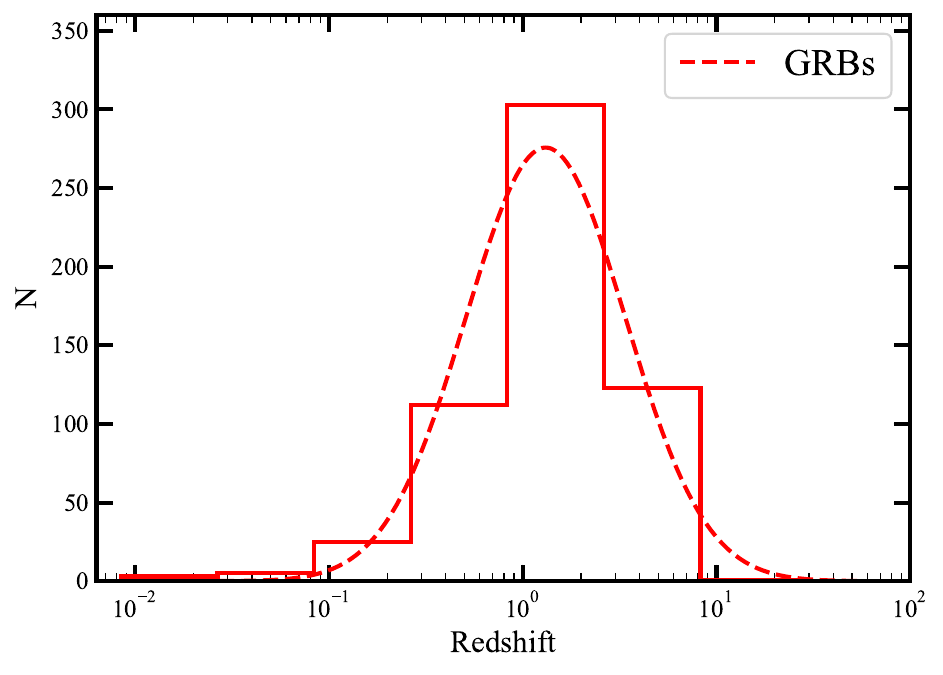}
	\includegraphics[angle=0,scale=0.50]{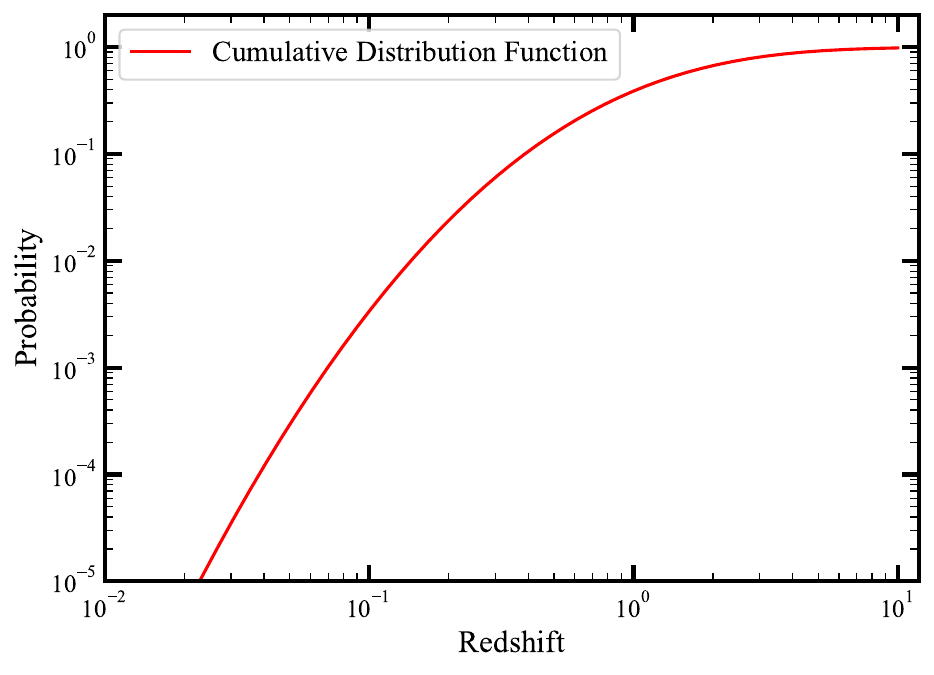}
	\caption{The redshift distributions of all GRBs with known redshift (left) and their cumulative distribution function (right). The red dotted line represents the best fitting with a lognormal function.}
	\label{figure:redshift}
\end{figure*}

\begin{figure*}
	\centering
	\includegraphics[angle=0,scale=0.80]{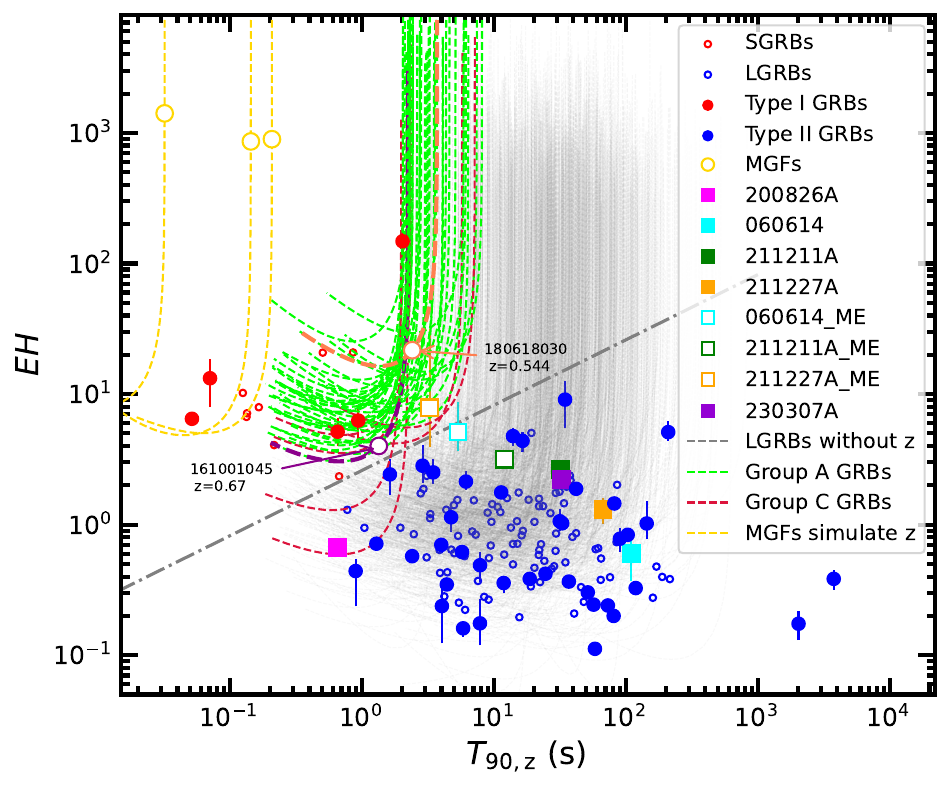}
	\includegraphics[angle=0,scale=0.80]{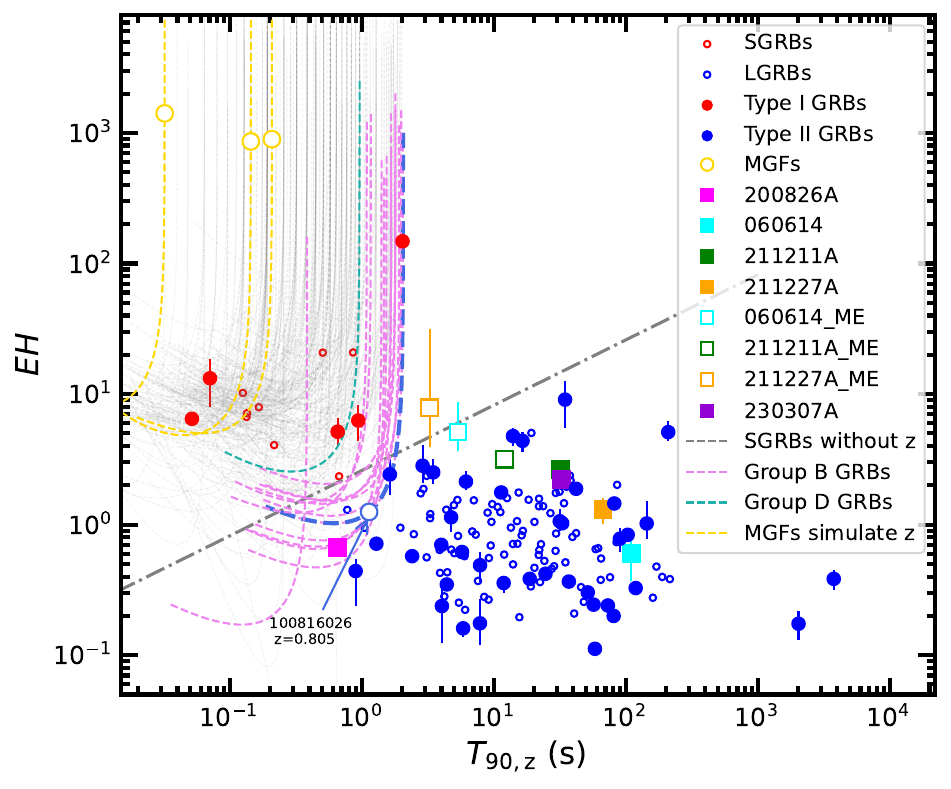}
	\caption{The $EH$--$T_{\rm 90,z}$ plane. The gray dotted line is $EHD=2.6$. The redshift increases gradually from the top right to the bottom left.}
	\label{figure:eh}
\end{figure*}

\subsubsection{$\varepsilon$ and $EHD$ parameter} \label{subsubsec:varepsilon}
According to the significant separation in the $E_{\rm p,z}$--$E_{\rm iso}$ plane, \cite{2010ApJ...725.1965L} and \cite{2020MNRAS.492.1919M} proposed two similar parameters to classify GRBs, $\varepsilon = (E_{\rm iso}/10^{52})/(E_{\rm p,z}/100)^{5/3}$ and the energy-hardness $EH = (E_{\rm p,z}/100)/(E_{\rm iso}/10^{51})^{0.4}$, respectively.
\cite{2010ApJ...725.1965L} suggested that the Type I GRBs have $\varepsilon<0.03$ and the rest frame duration $T_{90,z}<5$ s.
Subsequently, \cite{2020MNRAS.492.1919M} found a boundary in the $EH$--$T_{90}$ plane, and proposed a new parameter, the energy-hardness-duration $EHD = EH/T_{\rm 90,z}^{0.5}$, to classify GRBs with $EHD=2.6$.

We calculated $\varepsilon$ for peculiar GRBs at redshifts from 0.0001 to 10, as shown in Figure \ref{figure:epsilon}.
Group A GRBs and Group B GRBs are significantly different from the trajectories of LGRBs and SGRBs in the $\varepsilon$--$T_{\rm 90,z}$ plane.
In order to quantitatively study the classifications of peculiar GRBs, we obtain the redshift distribution of all GRBs with known redshifts and their cumulative distribution function (CDF), as shown in Figure \ref{figure:redshift}.
The redshifts of GRBs conform to a log-normal distribution with a median value of 1.31 and a dispersion of 0.41.
Next, we calculate the redshift range ($\rm z_1$-$\rm z_2$) of GRBs with $\varepsilon>0.03$, where $\rm z_1$ is the redshift corresponding to the $\varepsilon>0.03$ for the first time as the redshift increases from 0.0001, and $\rm z_2$ is the redshift corresponding to the $\varepsilon<0.03$ for the first time as the redshift increases from $\rm z_1$ to 10.
Using the CDF, we then calculate the probability of redshift less than $\rm z_1$ and $\rm z_2$, $P(\rm z < z_1)$ and $P(\rm  z<z_2)$, respectively.
The probability that the redshift of a GRB is within the redshift range making $\varepsilon>0.03$ is $P_\varepsilon=P(\rm z<z_2) - P(\rm z<z_1)$, which are listed in Table \ref{table:paticular}.
Note that, if $\varepsilon$ is never larger than 0.03, then the $P_\varepsilon=0$, and if $\varepsilon$ is not less than 0.03 as the redshift increases from $\rm z_1$ to 10, then $\rm z_2=10$.

The values of $\varepsilon$ of most Group A GRBs are less than 0.03 regardless of their redshift.
Among them, GRB 090518080, GRB 100719825, GRB 150228845, GRB 161001045, and GRB 161210524 crossed the boundary in a very small redshift range, with a probability of less than 50\%, which indicates that they are classified as Type I GRBs.
We calculated $\varepsilon$ of GRB 161001045 and GRB 180618030, $\varepsilon=0.014$ and $\varepsilon=0.001$ respectively, which are classified as Type I GRBs.
Group B GRBs are all classified as Type II GRBs within a redshift range of more than 60\% probability.
GRB 100816026 has $\varepsilon=0.124$ and is classified as Type II GRB.
For Group C GRBs, GRB 090320045 and GRB 101002279 with $\varepsilon$ consistently less than 0.03 are classified as Type I GRBs.
GRB 120504945 and GRB 140912664 with $\varepsilon$ over 0.03 in the small redshift range are also classified as Type I GRBs.
However, GRB 081130212 and GRB 131128629 with $\varepsilon$ over 0.03 in the large redshift range are classified as Type II GRBs.
Group D GRBs are classified as Type II GRBs in the redshift range with 73.8\% probability.

We calculated the $EH$ and $EHD$ for peculiar GRBs at redshifts from 0.0001 to 10, as shown in Figure \ref{figure:eh}.
We also calculated the redshift range of GRBs with $EHD< 2.6$ and calculate the probability that peculiar GRBs are within this redshift range using CDF, which is consistent with the method above.
The results are listed in Table \ref{table:paticular}.
Group A GRBs and Group B GRBs are significantly different from the trajectories of LGRBs and SGRBs in the $EH$--$T_{\rm 90,z}$ plane.
Group A and D GRBs are all classified as Type I GRBs within a redshift range.
GRB 161001045 and GRB 180618030 have $EHD$ = 3.46 and $EHD$ = 13.98, respectively, and are classified as Type I GRBs.
Group B GRBs are classified as Type II GRBs except for GRB 140626843 which has a 39.3\% probability in the redshift range of being classified as Type I GRBs.
GRB 100816026 has $EHD$ = 1.17 and is classified as Type II GRB.
For Group C GRBs, GRB 090320045, GRB 101002279, GRB 120504945, and GRB 140912664 with $EH$ consistently larger than 2.6 are classified as Type I GRBs.
GRB 081130212 and GRB 131128629 have larger than 90\% probability of being in the redshift range of being classified as Type II GRBs.

\begin{figure*}
	\centering
	\includegraphics[angle=0,scale=0.5]{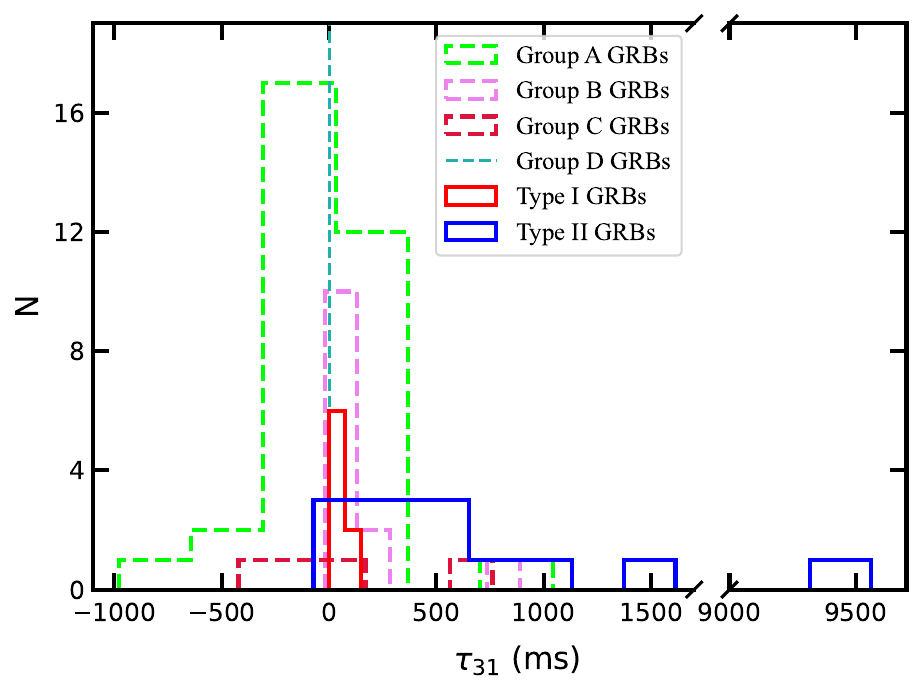}
	\caption{The $\tau_{31}$ distributions of Group A, B, C, and D GRBs, Type I GRBs and Type II GRBs. }
	\label{figure:tau}
\end{figure*}

\begin{figure*}
	\centering
	\includegraphics[angle=0,scale=0.8]{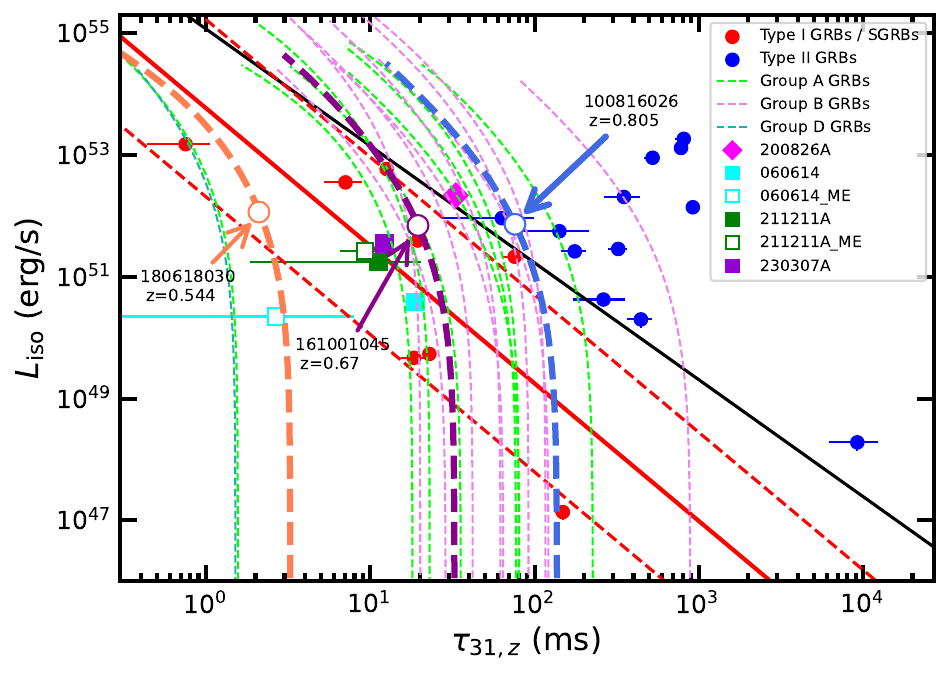}
	\caption{The $L_{\rm iso}$--$\tau_{\rm 31,z}$ plane. The black soild line is the possible dividing line. The red soild line is the best-fit line for Type I GRBs/SGRBs. The red dashed lines are the confidence regions of 2$\sigma$. The redshift increases gradually from the bottom right to the top left.}
	\label{figure:tau-liso}
\end{figure*}
\subsection{Spectral Lag} \label{subsec:lag}
Spectral lag has been widely used to provide clues to the progenitors of GRBs.
In addition, \cite{2000ApJ...534..248N} found that the isotropic luminosity--spectral lag ($L_{\rm iso}$--$\tau$) correlation in the LGRBs, except for GRB 980425 that is a low-luminosity GRB and is associated with SN 1998bw.
\cite{2012MNRAS.419..614U} further confirmed the $L_{\rm iso}$--$\tau$ correlation by using $\tau$ extracted in the rest frame.
However, \cite{2015MNRAS.446.1129B} raised concerns about the use of spectral lag to classify the progenitors of GRBs, as LGRBs do not always have significant spectral lag, which can significantly weaken the $L_{\rm iso}$--$\tau$ correlation.
They believed that the $L_{\rm iso}$--$\tau$ correlation may only be a boundary.
\cite{2023MNRAS.524.1096L} found that Type I GRBs and Type II GRBs follow different $L_{\rm iso}$--$\tau$ correlation, which may enable the identification of progenitors of GRBs.
However, some Type II GRBs with high luminosity or negative spectral lag do not follow the $L_{\rm iso}$--$\tau$ correlation of Type II GRBs.
Although spectral lag can not provide conclusive evidence for identifying the progenitors of GRBs, it still provides key clues.

We use the cross-correlation function (CCF) to calculate the $\tau$ of lightcurves among different energy bands.
For two lightcurves $x_{i}$ and $y_{i}$, CCF is defined as \citep{1997ApJ...486..928B}
\begin{equation}\label{ccf}
	{\rm CCF}_{\rm Band} (d, x, y)=\frac{ {\textstyle \sum_{i={\rm max}(1, 1-d)}^{{\rm min}(N, N-d)}}x_{i}y_{i+d} }{\sqrt{ {\textstyle \sum_{i}x_{i}^{2} {\textstyle \sum_{i}}y_{i}^{2} } } } ,
\end{equation}
where N is the total number of time bin ($t_{\rm bin}$), $d$ is the time delay in unit a $t_{\rm bin}$ .
We fitted the CCF around the peak with a polynomial to find $d_{\rm max}$ corresponding to its global maximum and calculated the spectral lag by $\tau= d \times t_{\rm bin}$.
The uncertainty of $\tau$ is estimated by Monte Carlo simulation \citep{2010ApJ...711.1073U}.
In our analysis, we estimate the $\tau$ of 10--25 keV with respect to 50-100 keV ($\tau_{31}$), the results are listed in Table \ref{table:paticular}.

The $\tau_{31}$ distributions and the $L_{\rm iso}$--$\tau_{\rm 31,z}$ plane are shown in Figure \ref{figure:tau} and Figure \ref{figure:tau-liso}, respectively, where $\tau_{\rm 31,z} = \tau_{\rm 31}/(1+z)$.
We find no significant difference in the distribution of $\tau$ between different groups nor between them and Type I GRBs and Type II GRBs.
However, 14 Group A GRBs, accounting for about 34\%, have negative $\tau$, which suggests that they are more likely to be SGRBs than LGRBs.
When only GRBs with positive $\tau_{\rm 31,z}$ are considered, there is a weak correlation $L_{\rm iso} \propto \tau_{\rm 31,z}^{-2.45}$ between $L_{\rm iso}$ and $\tau_{\rm 31,z}$ of both Type I GRBs and SGRBs, with a Pearsons coefficient $r = -0.75$, whereas Type II GRBs are not.
Although the $L_{\rm iso}$--$\tau_{\rm 31,z}$ correlation does not exist in Type II GRBs, it seems to serve as a dividing line to classify the different progenitors of GRBs.
If, as we expect, Group A GRBs have smaller redshifts and Group B GRBs have larger redshifts, this will support the notion that mergers are origin of Group A GRBs and collapsars are origin of Group B GRBs.
However, counter examples do exist - $\tau=150 \pm 15$ ms of GRB 170817A is significantly larger than that of GRB 200826A, and $\tau$ of GRB 180728A is even negative. This suggests that using $\tau$ alone as the only indicator cannot reliably differentiate progenitors.

\section{Discussion} \label{sec:discussions}

All Group A GRBs are classified as Type I GRBs in both the $\varepsilon$--$T_{\rm 90,z}$ and the $EH$--$T_{\rm 90,z}$ planes.
10 Group A GRBs are classified as LGRBs in the $E_{\rm p}$--$T_{90}$ plane or the $HR$--$T_{90}$ plane, and only GRB 081006604 is classified as LGRBs at the same time.
GRB 161001045 and GRB 180618030 have z = 0.67 and z = 0.554, respectively, and are classified as Type I GRBs in both the $\varepsilon$--$T_{\rm 90,z}$ and the $EH$--$T_{\rm 90,z}$ planes.
Moreover, according to the two type progenitors of GRBs, the properties of their host galaxies and GRB offset from the host galaxy center are significantly different \citep{2009ApJ...703.1696Z,2010ApJ...708....9F,2016ApJS..227....7L,2020ApJ...897..154L,2022ApJ...940...56F,2022ApJ...940...57N}. 
GRBs originating from mergers are usually located in the dwarf and elliptical galaxies, and have larger offset from the galactic center, their host galaxies also have smaller star formation rate (SFR) and smaller specific star formation rate (sSFR, sSFR = SFR/$M_{\odot}$, where $M_{\odot}$ is the stellar mass of the host galaxy), since the SN that needs to be experienced in the formation of an NS or BH kicks the progenitor away from the star-forming region of the host galaxy to bring about a larger offset \citep{2005Natur.438..988B,2005Natur.437..851G,2022ApJ...940...56F,2022ApJ...940...57N}.
However, GRBs originating from collapsars are usually located in the bright regions of the irregular and dwarf galaxies, where they exhibit a small offset and a large star formation rate.
GRB 161001045 is $18.54 \pm 6.22$ kpc away from the host galaxy, which is consistent with the merger origin \citep{2022ApJ...940...56F}.
Meanwhile, the host galaxy of GRB 161001045 with SFR = $0.53^{+0.59}_{-0.31}$ $M_{\odot}$ yr$^{-1}$ and log(sSFR) = $-10.02^{+0.33}_{-0.37}$ yr$^{-1}$ is a low-SFR galaxy. This is also consistent with the merger origin \citep{2022ApJ...940...57N}.
GRB 180618030 has an offset $9.7 \pm 1.69$ kpc from the host galaxy with an SFR = $1.85^{+1.77}_{-1.10}$ $M_{\odot}$ yr$^{-1}$ and log(sSFR) = $-8.54^{+0.44}_{-0.39}$ yr$^{-1}$, which is consistent with the merger origin \citep{2022ApJ...940...56F,2022ApJ...940...57N}.
In total, 17 Group A GRBs are suggested by all 5 indicators, including $\tau$, to be originated by mergers.

Among Group B GRBs, only GRB 140626843 is classified as Type I GRB in the $EH$--$T_{\rm 90,z}$ plane, and all other GRBs are classified as Type II GRBs in both the $\varepsilon$--$T_{\rm 90,z}$ and the $EH$--$T_{\rm 90,z}$ planes.
However, the extremely large $\tau$ of GRB 140626843 strongly supports its collapsar origin.
Despite the fact that classifications in both the $E_{\rm p}$--$T_{90}$ and the $HR$--$T_{90}$ planes do not support their collapsar origins while considering that GRB 200826A would be equally unsupported as a collapsar origin, we nevertheless argue that they should be collapsar origins.

Among Group C GRBs, GRB 081130212 and GRB 131128629 are classified as Type II GRBs, and other GRBs are classified as Type I GRBs in both the $\varepsilon$--$T_{\rm 90,z}$ and the $EH$--$T_{\rm 90,z}$ planes.
GRB 090320045 and GRB 140912664 are classified as Type I GRBs and other GRBs are classified as Type II GRBs in the $E_{\rm p}$--$T_{90}$ plane.
GRB 101002279 and GRB 12050494 are classified as Type II GRBs and other GRBs are classified as Type I GRBs in the $HR$--$T_{90}$ plane.
Just merger origins of GRB 090320045 and GRB 140912664 are supported by 4 clues.

Among Group D GRBs, GRB 150819440 is classified as Type II GRB in the $\varepsilon$--$T_{\rm 90,z}$ plane and is classified as Type I GRB in the other planes, therefore we can not determine its origin.

We compared the trajectories of MGFs with SGRBs in the $E_{\rm p,z}$--$E_{\rm iso}$, the $\varepsilon$--$T_{\rm 90,z}$, the $EH$--$T_{\rm 90,z}$ planes and found that they are not significantly different from SGRBs, and thus only precise localization is possible for identification.
Since all known durations of MGFs are significantly smaller than those of GRBs in our sample, they are not considered as being MGFs in this work.

\section{Conclusions} \label{sec:conclusions}

The identification of progenitors of GRBs is crucial for studying the physical mechanisms of GRBs.
Currently, convincing evidence to a certain type of progenitors comes from optical afterglow, with KNe pointing to mergers and SNe to collapsars.
However, KN/SN is dependent on high-precision multiband observations, which are bound to be lacking for the vast majority of GRBs.
The prompt emissions of GRBs also provide key clues to identify their progenitors, such as $T_{90}$, $E_{\rm p}$, and $\tau$, especially when their redshift can be detected.
$T_{90}$ has been used as an excellent alternative because it can be extracted in the lightcurves of almost all GRBs.
The earlier observed KNe/SNe corresponds perfectly with the $T_{90}$ classification.
However, GRB 200826A and GRB 211211A break the correspondence between the progenitors and $T_{90}$.

In this work, we attempted to determine the progenitors of peculiar GRBs that are identified through machine learning.
The peculiar GRBs are divided into four groups, depending on the machine learning classification results.
In order to verify whether long GRBs-I and short GRBs-II originated from mergers and collapsars, respectively, we investigated them from the $E_{\rm p}$--$T_{90}$, the $HR$--$T_{90}$, the $E_{\rm p,z}$--$E_{\rm iso}$, the $\varepsilon$--$T_{\rm 90,z}$, the $EH$--$T_{\rm 90,z}$, and the $L_{\rm iso}$--$\tau_{\rm 31,z}$ planes.

Finally, we found 17 long Type I GRB candidates and 13 short Type II GRB candidates, greatly expanding the sample of peculiar GRBs. As a result of the analysis above, our conclusions are as follows:

\begin{enumerate}
	\item There are 17 Group A GRBs, including GRB 161001045 and GRB 180618030, which are LGRBs but their $\tau$ are smaller than typical LGRBs.
	Meanwhile, they have a less than 50\% probability of being classified as LGRBs in both the $E_{\rm p}$--$T_{90}$ and the $HR$--$T_{90}$ planes.
	Moreover, in the $E_{\rm p,z}$--$E_{\rm iso}$, the $\varepsilon$--$T_{\rm 90,z}$, and the $EH$--$T_{\rm 90,z}$ planes, they demonstrate distinct properties compared to LGRBs and Type II GRBs.
	The host galaxy characteristics of both GRB 161001045 and GRB 180618030 are also consistent with a merger scenario.
	Although their $T_{90}$ are shorter than WE durations of long Type I GRBs, they are similar to ME durations of long Type I GRBs.
	These are strong evidence supporting their merger origin.
	Unfortunately, multiwavelength observations are lacking. 
	Assuming that Group A GRBs are indeed originated from mergers, this suggests that the population of long Type I GRBs is not scarce and that the lack of high-precision observations has prevented us from identifying them.
	Meanwhile, the timescale of compact binary mergers can extend to the same timescales of MEs of long Type I GRBs, which implies that EEs of long Type I GRBs may involve a distinct physical mechanism compared to MEs.

	\item Similar to GRB 200826A, GRB 100816026 has a $\tau$ larger than typical SGRBs and is consistent with LGRBs and Type II GRBs in the $E_{\rm p,z}$--$E_{\rm iso}$, $\varepsilon$--$T_{\rm 90,z}$ and the $EH$--$T_{\rm 90,z}$ planes, so it should originate from a collapsar.
	Although the absence of redshift does not allow the other Group B GRBs to be unambiguously localized in the $E_{\rm p,z}$--$E_{\rm iso}$, the $\varepsilon$--$T_{\rm 90,z}$, and the $EH$--$T_{\rm 90,z}$ planes, the probability of them belonging to Type II GRBs given by the distribution of redshifts strongly supports their collapsar origin.
	Assuming that Group B GRBs indeed originated from collapsars, this suggests that the population of long Type II GRBs is not scarce, which is unfavourable to the interpretation that the jet progression leading to a missed ME.
	
	\item The origins of Group C GRBs are mixed, GRB 081130212 and GRB 131128629 are likely to have collapsar origin but GRB 090320045, GRB 101002279, and GRB 140912664 are likely to have merger origin.
	
	\item GRB 150819440 is classified as Type II GRB in the $\varepsilon$--$T_{\rm 90,z}$ plane and is classified as Type I GRBs in the other planes.
	The origin of Group D is also ambiguous.
\end{enumerate}

\acknowledgments
We thank the anonymous reviewers for their insightful comments/suggestions.
We acknowledge the use of public data and software provided by the Fermi Science Support Center.
This work was supported in part by the National Natural Science Foundation of China (No. 12273122).


\begin{longrotatetable}
	\begin{deluxetable*}{lcccc ccccc ccccc }
		\tablecaption{The prompt emission parameters of GRBs associated with SN or GW/KN observations in the sample}
		\label{table:rest}
		\tabletypesize{\tiny}
		\tablehead{
			\colhead{$GRB$} & \colhead{$z$} & \colhead{$T_{90}$} & \colhead{$\tau_{31}$} & \colhead{$model$} & \colhead{$E_{\rm p}$} & \colhead{$\alpha$} &
			\colhead{$\beta$} & \colhead{$S_{\gamma,-6}$} & \colhead{$F_{\rm p,-6}$} & \colhead{$E_{\rm range}$} & \colhead{$E_{\rm iso,50}$} & \colhead{$L_{\rm iso,50}$} & \colhead{Ass.} & \colhead{Ref.}	\\
			\colhead{} & \colhead{} & \colhead{(s)} & \colhead{(ms)} & \colhead{} & \colhead{(keV)} & \colhead{} & \colhead{} & \colhead{(erg cm$^{-2}$)} & \colhead{(erg cm$^{-2}$ s$^{-1}$)} & \colhead{} & \colhead{(erg)} & \colhead{(erg s$^{-1}$)} & \colhead{} & \colhead{}
		}
		\startdata
		970228 & 0.695 & $53.44 \pm 2.89$ & -- & CPL & $165^{+39}_{-25}$ & -1.27 &  -- & $8.07 \pm 0.49$ & $4.59 \pm 0.8$ & 10--10000 & $113^{+6.84}_{-6.84}$ & $109^{+18.9}_{-18.9}$ & SN & (1) \\
		980326 & 0.9 & $9 \pm 0.9$ & -- & Band & $92^{+20}_{-20}$ & -1.23 & -2.48 & $0.76 \pm 0.15$ & $0.25 \pm 0.02$ & 40--700 & $29.2^{+5.77}_{-5.77}$ & $17.9^{+1.1}_{-1.1}$ & SN & (2) \\
		980425 & 0.0085 & $34.88 \pm 3.78$ & -- & Band & $54.6^{+20.9}_{-20.9}$ & -1 & -2.1 & $4 \pm 0.74$ & $0.29 \pm 0.02$ & 20--2000 & $0.0091^{+0.0017}_{-0.0017}$ & $0.0007^{+0.00004}_{-0.00004}$ & SN & (3),(4) \\
		990712 & 0.434 & $20 \pm 2$ & -- & Band & $779^{+125}_{-125}$ & -1.88 & -2.48 & $6.5 \pm 0.3$ & $1.3 \pm 0.1$ & 40--700 & $84.2^{+3.89}_{-3.89}$ & $24.1^{+1.86}_{-1.86}$ & SN & (2) \\
		991208 & 0.706 & $63.06 \pm 0.48$ & -- & Band & $185^{+5}_{-5}$ & -1.27 & -2.92 & $155 \pm 2.9$ & $22 \pm 1.1$ & 10--10000 & $2180^{+40.7}_{-40.7}$ & $527^{+26.4}_{-26.4}$ & SN & (1) \\
		000911 & 1.058 & $23.35 \pm 0.23$ & -- & Band & $1083^{+52}_{-50}$ & -0.82 & -2.75 & $208 \pm 6.5$ & $29 \pm 2.8$ & 10--10000 & $5680^{+177}_{-177}$ & $1630^{+157}_{-157}$ & SN & (1) \\
		011121 & 0.36 & $57.01 \pm 3.78$ & -- & Band & $819^{+108}_{-96}$ & -1.26 & -2.01 & $279 \pm 7.5$ & $28.2 \pm 2.3$ & 10--10000 & $852^{+22.9}_{-22.9}$ & $117^{+9.56}_{-9.56}$ & SN & (1) \\
		020405 & 0.69 & $41.54 \pm 2.03$ & -- & Band & $161^{+7}_{-7}$ & -0.97 & -2.6 & $83.7 \pm 1.2$ & $8.16 \pm 0.93$ & 10--10000 & $1060^{+15.3}_{-15.3}$ & $175^{+20}_{-20}$ & SN & (1) \\
		020903 & 0.251 & $9.8 \pm 0.6$ & -- & CPL & $2.6^{+1.4}_{-0.8}$ & -1 &  -- & $0.06 \pm 0.01$ & $0.01 \pm 0.01$ & 2--10 & $0.15^{+0.04}_{-0.04}$ & $0.05^{+0.02}_{-0.02}$ & -- & (5) \\
		021211 & 1.01 & $1.8 \pm 0.25$ & -- & CPL & $54^{+13}_{-25}$ & -1.43 &  -- & $2.62 \pm 0.26$ & $2.62 \pm 0.4$ & 10--10000 & $95^{+9.43}_{-9.43}$ & $191^{+29.2}_{-29.2}$ & SN & (1) \\
		030329 & 0.168 & $21.81 \pm 1.15$ & -- & Band & $97^{+2}_{-2}$ & -1.53 & -2.78 & $187 \pm 2.6$ & $24.7 \pm 1.4$ & 10--10000 & $149^{+2.07}_{-2.07}$ & $23^{+1.3}_{-1.3}$ & SN & (1) \\
		040924 & 0.859 & $2.39 \pm 0.24$ & -- & CPL & $72^{+6}_{-6}$ & -0.55 &  -- & $2.31 \pm 0.07$ & $3.99 \pm 0.5$ & 10--10000 & $48.1^{+1.35}_{-1.35}$ & $154^{+19.3}_{-19.3}$ & SN & (1),(6) \\
		041006 & 0.716 & $6.85 \pm 1.98$ & -- & CPL & $86^{+5}_{-5}$ & 0.08 &  -- & $4.72 \pm 0.25$ & $2.27 \pm 0.48$ & 10--10000 & $64.5^{+3.42}_{-3.42}$ & $53.3^{+11.3}_{-11.3}$ & SN & (1) \\
		050416A & 0.6535 & $6.67 \pm 3.42$ & -- & CPL & $14.85^{+7.13}_{-7.13}$ & -0.82 &  -- & $0.37 \pm 0.05$ & $0.19 \pm 0.02$ & 15--150 & $10.8^{+1.46}_{-1.46}$ & $9.12^{+0.9}_{-0.9}$ & -- & (7) \\
		050525A & 0.606 & $7.05 \pm 0.19$ & $562.01 \pm 138.09$ & Band & $80^{+2}_{-2}$ & -1.05 & -3.2 & $24.6 \pm 0.44$ & $12 \pm 0.73$ & 10--10000 & $262^{+4.68}_{-4.68}$ & $205^{+12.5}_{-12.5}$ & SN & (1),(8) \\
		060218 & 0.0331 & $2100 \pm 100$ & -- & Band & $4.67^{+1.12}_{-1.15}$ & -1.44 & -2.54 & $17.2 \pm 1.8$ & $0.0011 \pm 0.0005$ & 0.5--150 & $0.4^{+0.04}_{-0.04}$ & $0.00003^{+0.00001}_{-0.00001}$ & -- & (9) \\
		060614 & 0.125 & $123.65 \pm 18.94$ & $21.2 \pm 2.86$ & CPL & $76^{+20}_{-29}$ & -1.92 &  -- & $47 \pm 2.6$ & $6.6 \pm 1.3$ & 10--10000 & $24.4^{+1.35}_{-1.35}$ & $3.86^{+0.76}_{-0.76}$ & KN & (10) \\
		060614$^{*}$ & 0.125 & $6 \pm 0.6$ & $3 \pm 6$ & CPL & $302^{+214}_{-85}$ & -1.57 &  -- & $8.19 \pm 0.56$ & $4.5 \pm 0.72$ & 20--20000 & $3.6^{+0.25}_{-0.25}$ & $2.23^{+0.36}_{-0.36}$ & KN & (10),(11) \\
		070714B & 0.92 & $1.26 \pm 0.17$ & $24.14 \pm 2.31$ & CPL & $551^{+148}_{-112}$ & 0.19 &  -- & $2.71 \pm 0.54$ & $13.8 \pm 3.1$ & 10--10000 & $60.5^{+12.1}_{-12.1}$ & $591^{+133}_{-133}$ & KN & (1) \\
		071112C & 0.823 & $5.25 \pm 0.83$ & -- & CPL & $406^{+179}_{-100}$ & -1.13 &  -- & $6.02 \pm 1.11$ & $3.8 \pm 1.1$ & 10--10000 & $111^{+20.5}_{-20.5}$ & $128^{+37}_{-37}$ & SN & (1) \\
		080319B & 1.95 & $10.23 \pm 0.64$ & $2298.46 \pm 59.46$ & CPL & $632^{+160}_{-113}$ & -1.21 &  -- & $15.5 \pm 1.8$ & $4.61 \pm 0.83$ & 10--10000 & $1500^{+174}_{-174}$ & $1310^{+237}_{-237}$ & SN & (1) \\
		080905 & 0.1218 & $0.96 \pm 0.35$ & -- & CPL & $349.71^{+55.27}_{-55.27}$ & 0.2 &  -- & $0.45 \pm 0.06$ & $1.97 \pm 0.21$ & 10--1000 & $0.15^{+0.02}_{-0.02}$ & $0.76^{+0.08}_{-0.08}$ & -- & (12) \\
		080905B & 2.374 & $105.98 \pm 6.8$ & -- & CPL & $199.14^{+31.41}_{-31.41}$ & -0.88 &  -- & $1.96 \pm 0.22$ & $0.53 \pm 0.14$ & 10--1000 & $274^{+30.7}_{-30.7}$ & $249^{+67.1}_{-67.1}$ & -- & (12) \\
		080916 & 0.689 & $46.34 \pm 7.17$ & -- & Band & $105.73^{+20.45}_{-20.45}$ & -0.78 & -1.77 & $12.8 \pm 0.4$ & $1.05 \pm 0.2$ & 10--1000 & $299^{+9.22}_{-9.22}$ & $41.3^{+7.87}_{-7.87}$ & -- & (12) \\
		081007 & 0.5295 & $12 \pm 0.12$ & $401.73 \pm 140.92$ & CPL & $40^{+10}_{-10}$ & -1.4 &  -- & $1.2 \pm 0.1$ & $0.19 \pm 0.02$ & 25--900 & $17.5^{+1.46}_{-1.46}$ & $4.24^{+0.39}_{-0.39}$ & SN & (13),(8) \\
		081121 & 2.512 & $41.99 \pm 8.51$ & -- & Band & $160.94^{+14.45}_{-14.45}$ & -0.44 & -2.1 & $14.8 \pm 0.42$ & $2.19 \pm 0.28$ & 10--1000 & $2850^{+80.7}_{-80.7}$ & $1480^{+192}_{-192}$ & -- & (12) \\
		081221 & 2.26 & $29.7 \pm 0.41$ & -- & CPL & $88.4^{+1.11}_{-1.11}$ & -0.86 &  -- & $28.7 \pm 0.23$ & $2.08 \pm 0.1$ & 10--1000 & $3850^{+31.1}_{-31.1}$ & $908^{+45}_{-45}$ & -- & (12) \\
		081222 & 2.77 & $18.88 \pm 2.32$ & -- & Band & $147.21^{+8.43}_{-8.43}$ & -0.84 & -2.3 & $12.1 \pm 0.34$ & $1.85 \pm 0.13$ & 10--1000 & $2550^{+72.6}_{-72.6}$ & $1480^{+102}_{-102}$ & -- & (12) \\
		090323 & 3.57 & $133.89 \pm 0.57$ & -- & SBPL & $397.51^{+62.33}_{-62.33}$ & -1.24 & -2.13 & $128.67 \pm 1.2$ & $2.69 \pm 0.2$ & 10--1000 & $44600^{+416}_{-416}$ & $4250^{+321}_{-321}$ & -- & (12) \\
		090328 & 0.736 & $61.7 \pm 1.81$ & -- & CPL & $711.34^{+39.28}_{-39.28}$ & -1.1 &  -- & $52.5 \pm 0.6$ & $5.41 \pm 0.32$ & 10--1000 & $1010^{+11.5}_{-11.5}$ & $181^{+10.7}_{-10.7}$ & -- & (12) \\
		090423 & 8.26 & $7.17 \pm 2.42$ & -- & CPL & $71.49^{+8.7}_{-8.7}$ & -0.65 &  -- & $0.57 \pm 0.06$ & $0.29 \pm 0.08$ & 10--1000 & $588^{+58.9}_{-58.9}$ & $2730^{+788}_{-788}$ & -- & (12) \\
		090424 & 0.544 & $14.14 \pm 0.26$ & -- & Band & $160^{+3.98}_{-3.98}$ & -1.02 & -2.76 & $47 \pm 0.53$ & $14.32 \pm 0.23$ & 10--1000 & $418^{+4.74}_{-4.74}$ & $196^{+3.17}_{-3.17}$ & -- & (12) \\
		090510 & 0.903 & $0.96 \pm 0.14$ & -- & CPL & $4727.06^{+348.98}_{-348.98}$ & -0.86 &  -- & $4.42 \pm 0.11$ & $15.81 \pm 0.61$ & 10--1000 & $391^{+9.28}_{-9.28}$ & $2660^{+103}_{-103}$ & -- & (12) \\
		090516 & 4.109 & $123.14 \pm 2.06$ & -- & CPL & $159.36^{+15.51}_{-15.51}$ & -1.47 &  -- & $23.1 \pm 0.91$ & $0.69 \pm 0.13$ & 10--1000 & $9350^{+367}_{-367}$ & $1430^{+279}_{-279}$ & -- & (12) \\
		090618 & 0.54 & $112.39 \pm 1.09$ & $1415.31 \pm 118.45$ & Band & $149.04^{+3.29}_{-3.29}$ & -1.11 & -2.24 & $274.41 \pm 1.51$ & $8.75 \pm 0.55$ & 10--1000 & $2820^{+15.5}_{-15.5}$ & $139^{+8.66}_{-8.66}$ & -- & (12) \\
		090902B & 1.822 & $19.33 \pm 0.29$ & -- & SBPL & $789.1^{+59.14}_{-59.14}$ & -1.09 & -4.86 & $279.38 \pm 1.13$ & $27.13 \pm 0.52$ & 10--1000 & $31700^{+128}_{-128}$ & $8680^{+166}_{-166}$ & -- & (12) \\
		090926 & 2.1062 & $13.76 \pm 0.29$ & -- & SBPL & $301.39^{+12.59}_{-12.59}$ & -0.98 & -2.31 & $152.89 \pm 0.68$ & $25.83 \pm 0.38$ & 10--1000 & $21900^{+97.5}_{-97.5}$ & $11500^{+170}_{-170}$ & -- & (12) \\
		090926B & 1.24 & $64 \pm 1.56$ & -- & CPL & $84.8^{+2.05}_{-2.05}$ & 0.16 &  -- & $7.57 \pm 0.17$ & $0.53 \pm 0.11$ & 10--1000 & $311^{+6.9}_{-6.9}$ & $48.4^{+10.5}_{-10.5}$ & -- & (12) \\
		090927 & 1.37 & $0.51 \pm 0.23$ & -- & CPL & $175.81^{+41.39}_{-41.39}$ & -0.64 &  -- & $0.21 \pm 0.04$ & $0.85 \pm 0.14$ & 10--1000 & $10.6^{+2.01}_{-2.01}$ & $101^{+16.8}_{-16.8}$ & -- & (12) \\
		091003 & 0.8969 & $20.22 \pm 0.36$ & -- & CPL & $432.21^{+21.1}_{-21.1}$ & -1.1 &  -- & $36 \pm 0.56$ & $8.19 \pm 0.39$ & 10--1000 & $882^{+13.7}_{-13.7}$ & $381^{+18.1}_{-18.1}$ & -- & (12) \\
		091020 & 1.71 & $24.26 \pm 7.97$ & -- & CPL & $243.84^{+27.47}_{-27.47}$ & -1.26 &  -- & $8.01 \pm 0.4$ & $1.24 \pm 0.15$ & 10--1000 & $662^{+33.3}_{-33.3}$ & $277^{+34.1}_{-34.1}$ & -- & (12) \\
		091024 & 1.092 & $450.57 \pm 2.36$ & -- & Band & $162.5^{+9.99}_{-9.99}$ & -0.84 & -2.15 & $54.1 \pm 1.52$ & $0.93 \pm 0.2$ & 10--1000 & $2350^{+66.1}_{-66.1}$ & $84.3^{+18.1}_{-18.1}$ & -- & (12) \\
		091127 & 0.49 & $8.7 \pm 0.57$ & $94.2 \pm 53.86$ & SBPL & $32.73^{+4.43}_{-4.43}$ & -1.29 & -2.23 & $18.3 \pm 0.21$ & $7.03 \pm 0.15$ & 10--1000 & $161^{+1.8}_{-1.8}$ & $92^{+1.98}_{-1.98}$ & -- & (12) \\
		091208B & 1.063 & $12.48 \pm 5.02$ & -- & Band & $44.73^{+12.76}_{-12.76}$ & -0.62 & -1.92 & $7.69 \pm 0.29$ & $3.16 \pm 0.15$ & 10--1000 & $350^{+13.3}_{-13.3}$ & $297^{+13.7}_{-13.7}$ & -- & (12) \\
		100117A & 0.92 & $0.26 \pm 0.83$ & -- & CPL & $325.43^{+51.09}_{-51.09}$ & -0.1 &  -- & $0.37 \pm 0.06$ & $2.1 \pm 0.23$ & 10--1000 & $8.46^{+1.29}_{-1.29}$ & $91.6^{+9.88}_{-9.88}$ & -- & (12) \\
		100206A & 0.4068 & $0.18 \pm 0.07$ & -- & CPL & $531.82^{+71.71}_{-71.71}$ & -0.4 &  -- & $0.97 \pm 0.06$ & $8.33 \pm 0.38$ & 10--1000 & $4.61^{+0.29}_{-0.29}$ & $55.7^{+2.56}_{-2.56}$ & -- & (12) \\
		100414A & 1.368 & $26.5 \pm 2.07$ & -- & CPL & $668.13^{+14.63}_{-14.63}$ & -0.63 &  -- & $91.9 \pm 0.66$ & $9.01 \pm 0.34$ & 10--1000 & $5740^{+41}_{-41}$ & $1330^{+49.6}_{-49.6}$ & -- & (12) \\
		100615A & 1.398 & $37.38 \pm 0.98$ & -- & Band & $53.55^{+7.5}_{-7.5}$ & -0.91 & -1.8 & $13.3 \pm 0.32$ & $1.16 \pm 0.11$ & 10--1000 & $1110^{+26.5}_{-26.5}$ & $232^{+22}_{-22}$ & -- & (12) \\
		100625A & 0.452 & $0.24 \pm 0.28$ & $28.27 \pm 4.19$ & CPL & $483.19^{+63.32}_{-63.32}$ & -0.59 &  -- & $1.27 \pm 0.1$ & $4.69 \pm 0.72$ & 10--1000 & $7.37^{+0.55}_{-0.55}$ & $39.5^{+6.06}_{-6.06}$ & -- & (12) \\
		100728A & 1.567 & $165.38 \pm 2.9$ & -- & SBPL & $232.06^{+13.34}_{-13.34}$ & -0.72 & -2.48 & $124.54 \pm 1.28$ & $2.57 \pm 0.24$ & 10--1000 & $9850^{+101}_{-101}$ & $521^{+47.8}_{-47.8}$ & -- & (12) \\
		100728B & 2.106 & $10.24 \pm 1.85$ & -- & CPL & $153.35^{+18.46}_{-18.46}$ & -1.01 &  -- & $2.68 \pm 0.19$ & $0.87 \pm 0.13$ & 10--1000 & $311^{+21.5}_{-21.5}$ & $313^{+47.2}_{-47.2}$ & -- & (12) \\
		100814A & 1.44 & $150.53 \pm 1.62$ & -- & CPL & $146.68^{+6.75}_{-6.75}$ & -0.4 &  -- & $12.5 \pm 0.43$ & $0.84 \pm 0.17$ & 10--1000 & $685^{+23.3}_{-23.3}$ & $112^{+22.8}_{-22.8}$ & -- & (12) \\
		100816A & 0.8049 & $2.05 \pm 0.23$ & $137.06 \pm 7.38$ & CPL & $140.07^{+6.07}_{-6.07}$ & -0.37 &  -- & $3.35 \pm 0.11$ & $2.34 \pm 0.13$ & 10--1000 & $58.1^{+1.91}_{-1.91}$ & $73.1^{+3.99}_{-3.99}$ & -- & (12),This work \\
		100906A & 1.727 & $110.59 \pm 2.83$ & -- & Band & $74.91^{+24.27}_{-24.27}$ & -0.93 & -1.86 & $26.1 \pm 0.69$ & $2.29 \pm 0.22$ & 10--1000 & $3000^{+78.8}_{-78.8}$ & $719^{+68.6}_{-68.6}$ & -- & (12) \\
		101213A & 0.414 & $45.06 \pm 1.95$ & -- & CPL & $342.95^{+30.26}_{-30.26}$ & -0.95 &  -- & $13.1 \pm 0.6$ & $1.37 \pm 0.36$ & 10--1000 & $60.8^{+2.79}_{-2.79}$ & $9.01^{+2.36}_{-2.36}$ & -- & (12) \\
		101219B & 0.55 & $51.01 \pm 1.78$ & -- & CPL & $82.6^{+4.61}_{-4.61}$ & 0.08 &  -- & $2.2 \pm 0.11$ & $0.26 \pm 0.07$ & 10--1000 & $17.5^{+0.9}_{-0.9}$ & $3.23^{+0.86}_{-0.86}$ & -- & (12) \\
		110106B & 0.618 & $35.52 \pm 3.61$ & -- & CPL & $130.71^{+14.49}_{-14.49}$ & -0.94 &  -- & $1.97 \pm 0.17$ & $0.45 \pm 0.12$ & 10--1000 & $20.8^{+1.77}_{-1.77}$ & $7.73^{+1.98}_{-1.98}$ & -- & (12) \\
		110213A & 1.46 & $34.31 \pm 1.64$ & -- & CPL & $112.57^{+11.82}_{-11.82}$ & -1.56 &  -- & $10.4 \pm 0.43$ & $1.64 \pm 0.18$ & 10--1000 & $737^{+30.6}_{-30.6}$ & $286^{+30.6}_{-30.6}$ & -- & (12) \\
		110402A & 0.854 & $35.65 \pm 1.46$ & -- & CPL & $932.89^{+278.31}_{-278.31}$ & -1.33 &  -- & $7.39 \pm 0.04$ & $4.83 \pm 0.41$ & 10--1000 & $218^{+13.2}_{-13.2}$ & $264^{+22.5}_{-22.5}$ & -- & (12) \\
		110731A & 2.83 & $7.49 \pm 0.57$ & -- & SBPL & $289.29^{+61.6}_{-61.6}$ & -1.04 & -2.93 & $22.3 \pm 0.38$ & $5.26 \pm 0.38$ & 10--1000 & $4680^{+80.2}_{-80.2}$ & $4230^{+307}_{-307}$ & -- & (12) \\
		110818A & 3.36 & $67.07 \pm 3.92$ & -- & CPL & $314.95^{+54.65}_{-54.65}$ & -1.19 &  -- & $6.3 \pm 0.45$ & $0.69 \pm 0.21$ & 10--1000 & $1700^{+122}_{-122}$ & $811^{+250}_{-250}$ & -- & (12) \\
		111107A & 2.893 & $12.03 \pm 0.92$ & -- & CPL & $263.74^{+76.84}_{-76.84}$ & -1.28 &  -- & $1.86 \pm 0.27$ & $0.6 \pm 0.23$ & 10--1000 & $394^{+57.6}_{-57.6}$ & $491^{+191}_{-191}$ & -- & (12) \\
		111117A & 2.211 & $0.43 \pm 0.08$ & $2.41 \pm 1$ & CPL & $543.62^{+102.96}_{-102.96}$ & -0.5 &  -- & $0.68 \pm 0.06$ & $3.42 \pm 0.33$ & 10--1000 & $93.9^{+8.29}_{-8.29}$ & $1510^{+147}_{-147}$ & -- & (12),(8) \\
		111209A & 0.677 & $6300 \pm 630$ & -- & CPL & $310^{+53}_{-53}$ & -1.31 &  -- & $486 \pm 61$ & $0.18 \pm 0.02$ & 20--1400 & $6730^{+845}_{-845}$ & $4.27^{+0.43}_{-0.43}$ & -- & (14) \\
		111228A & 0.714 & $99.84 \pm 2.11$ & $242.67 \pm 124.02$ & Band & $26.55^{+1.37}_{-1.37}$ & -1.58 & -2.45 & $15.5 \pm 0.38$ & $1.52 \pm 0.1$ & 10--1000 & $334^{+8.07}_{-8.07}$ & $56.2^{+3.54}_{-3.54}$ & -- & (12),(8) \\
		120119A & 1.728 & $55.3 \pm 6.23$ & -- & Band & $182.76^{+10.45}_{-10.45}$ & -0.95 & -2.37 & $39.4 \pm 0.55$ & $2.97 \pm 0.23$ & 10--1000 & $3720^{+52.3}_{-52.3}$ & $765^{+58.4}_{-58.4}$ & -- & (12) \\
		120326A & 1.798 & $11.78 \pm 1.81$ & -- & SBPL & $43.24^{+10.05}_{-10.05}$ & -0.92 & -2.4 & $3.84 \pm 0.17$ & $0.76 \pm 0.09$ & 10--1000 & $382^{+17.3}_{-17.3}$ & $212^{+25.3}_{-25.3}$ & -- & (12) \\
		120624B & 2.1974 & $271.36 \pm 4.58$ & -- & SBPL & $592.78^{+76.33}_{-76.33}$ & -1.02 & -2.22 & $196.15 \pm 1.54$ & $5.14 \pm 0.3$ & 10--1000 & $33800^{+266}_{-266}$ & $2840^{+168}_{-168}$ & -- & (12) \\
		120711A & 1.405 & $44.03 \pm 0.72$ & -- & Band & $1317.52^{+42.26}_{-42.26}$ & -0.98 & -2.8 & $197.58 \pm 1$ & $12.66 \pm 0.8$ & 10--1000 & $18800^{+95.2}_{-95.2}$ & $2900^{+182}_{-182}$ & -- & (12) \\
		120712A & 4.1745 & $22.53 \pm 5.43$ & -- & CPL & $171.67^{+11.94}_{-11.94}$ & -0.33 &  -- & $3.52 \pm 0.18$ & $0.77 \pm 0.16$ & 10--1000 & $1220^{+62.8}_{-62.8}$ & $1390^{+279}_{-279}$ & -- & (12) \\
		120716A & 2.486 & $226.05 \pm 1.06$ & -- & CPL & $131.6^{+5.29}_{-5.29}$ & -0.96 &  -- & $12.9 \pm 0.3$ & $1.5 \pm 0.14$ & 10--1000 & $2010^{+46.7}_{-46.7}$ & $814^{+78.3}_{-78.3}$ & -- & (12) \\
		120811C & 2.671 & $14.34 \pm 6.56$ & -- & CPL & $60.81^{+3.73}_{-3.73}$ & -0.93 &  -- & $3.29 \pm 0.15$ & $0.82 \pm 0.17$ & 10--1000 & $625^{+28.5}_{-28.5}$ & $571^{+118}_{-118}$ & -- & (12) \\
		120907A & 0.97 & $5.76 \pm 1.78$ & -- & CPL & $120.71^{+22.14}_{-22.14}$ & -0.79 &  -- & $0.69 \pm 0.11$ & $0.71 \pm 0.18$ & 10--1000 & $18.1^{+2.85}_{-2.85}$ & $36.7^{+9.44}_{-9.44}$ & -- & (12) \\
		120909A & 3.93 & $112.07 \pm 10.42$ & -- & Band & $199.65^{+24.28}_{-24.28}$ & -0.84 & -1.93 & $18.1 \pm 0.57$ & $0.73 \pm 0.17$ & 10--1000 & $7200^{+228}_{-228}$ & $1430^{+329}_{-329}$ & -- & (12) \\
		121128A & 2.2 & $17.34 \pm 0.92$ & -- & SBPL & $56.49^{+8.52}_{-8.52}$ & -0.91 & -2.48 & $9.95 \pm 0.33$ & $2.11 \pm 0.19$ & 10--1000 & $1370^{+46}_{-46}$ & $932^{+83.7}_{-83.7}$ & -- & (12) \\
		121211A & 1.023 & $5.63 \pm 1.72$ & -- & CPL & $98.04^{+10.34}_{-10.34}$ & -0.22 &  -- & $0.49 \pm 0.05$ & $0.39 \pm 0.1$ & 10--1000 & $14^{+1.42}_{-1.42}$ & $22.5^{+5.6}_{-5.6}$ & -- & (12) \\
		130215A & 0.597 & $143.75 \pm 13.03$ & -- & Band & $209.95^{+42.31}_{-42.31}$ & -1.06 & -1.61 & $17.6 \pm 0.43$ & $0.9 \pm 0.2$ & 10--1000 & $384^{+9.37}_{-9.37}$ & $31.4^{+6.99}_{-6.99}$ & -- & (12) \\
		130420A & 1.297 & $104.96 \pm 8.81$ & -- & CPL & $56.98^{+3.02}_{-3.02}$ & -1.12 &  -- & $5.76 \pm 0.21$ & $0.53 \pm 0.12$ & 10--1000 & $308^{+11.1}_{-11.1}$ & $65.3^{+14.9}_{-14.9}$ & -- & (12) \\
		130427A & 0.3399 & $138.24 \pm 3.24$ & $1084.86 \pm 123.23$ & Band & $824.99^{+5.45}_{-5.45}$ & -1.02 & -2.83 & $1411.72 \pm 1.82$ & $305.44 \pm 2.55$ & 10--1000 & $6370^{+8.22}_{-8.22}$ & $1850^{+15.4}_{-15.4}$ & -- & (12),(8) \\
		130518A & 2.488 & $48.58 \pm 0.92$ & -- & SBPL & $419.83^{+57.72}_{-57.72}$ & -0.97 & -2.11 & $95.5 \pm 0.73$ & $11.05 \pm 0.37$ & 10--1000 & $19300^{+147}_{-147}$ & $7780^{+264}_{-264}$ & -- & (12) \\
		130603B & 0.36 & $0.07 \pm 0.01$ & $9.6 \pm 2.49$ & CPL & $607^{+61}_{-52}$ & -0.67 &  -- & $5.72 \pm 0.29$ & $81.3 \pm 5.2$ & 10--10000 & $18.4^{+0.93}_{-0.93}$ & $356^{+22.8}_{-22.8}$ & KN & (1),(8) \\
		130610A & 2.092 & $21.76 \pm 1.64$ & -- & CPL & $284.99^{+61.61}_{-61.61}$ & -1.58 &  -- & $5.72 \pm 0.47$ & $0.53 \pm 0.16$ & 10--1000 & $791^{+65.4}_{-65.4}$ & $227^{+66.8}_{-66.8}$ & -- & (12) \\
		130612A & 2.006 & $7.42 \pm 6.19$ & -- & CPL & $57.73^{+10}_{-10}$ & -1.37 &  -- & $0.46 \pm 0.05$ & $0.25 \pm 0.07$ & 10--1000 & $60.5^{+6.08}_{-6.08}$ & $99.4^{+25.9}_{-25.9}$ & -- & (12) \\
		130702A & 0.145 & $58.88 \pm 6.19$ & -- & CPL & $15.17^{+0.28}_{-0.28}$ & 1.6 &  -- & $3.67 \pm 0.06$ & $0.43 \pm 0.12$ & 10--1000 & $2.49^{+0.04}_{-0.04}$ & $0.34^{+0.1}_{-0.1}$ & -- & (12) \\
		130831A & 0.4791 & $17.53 \pm 2.81$ & $477.57 \pm 62.02$ & CPL & $54^{+7}_{-9}$ & -1.64 &  -- & $8.93 \pm 0.4$ & $2.32 \pm 0.28$ & 10--10000 & $74.8^{+3.35}_{-3.35}$ & $28.7^{+3.47}_{-3.47}$ & SN & (1),(8) \\
		130925A & 0.347 & $6.4 \pm 2.43$ & -- & CPL & $23.75^{+2.54}_{-2.54}$ & -0.69 &  -- & $0.42 \pm 0.03$ & $0.21 \pm 0.06$ & 10--1000 & $1.74^{+0.11}_{-0.11}$ & $1.19^{+0.33}_{-0.33}$ & -- & (12) \\
		130925A & 0.347 & $215.56 \pm 1.81$ & -- & CPL & $84.55^{+1.61}_{-1.61}$ & -1.66 &  -- & $86.7 \pm 0.72$ & $0.96 \pm 0.11$ & 10--1000 & $348^{+2.88}_{-2.88}$ & $5.17^{+0.57}_{-0.57}$ & -- & (12) \\
		131004A & 0.717 & $1.15 \pm 0.59$ & $130 \pm 20$ & CPL & $117.91^{+24.21}_{-24.21}$ & -1.36 &  -- & $0.44 \pm 0.05$ & $0.79 \pm 0.14$ & 10--1000 & $6.93^{+0.81}_{-0.81}$ & $21.4^{+3.73}_{-3.73}$ & -- & (12),(15) \\
		131011A & 1.874 & $77.06 \pm 3$ & -- & CPL & $274.19^{+24.27}_{-24.27}$ & -0.96 &  -- & $11.3 \pm 0.48$ & $1.02 \pm 0.2$ & 10--1000 & $1050^{+44.9}_{-44.9}$ & $274^{+52.4}_{-52.4}$ & -- & (12) \\
		131105A & 1.686 & $112.64 \pm 0.46$ & -- & CPL & $266.11^{+15.58}_{-15.58}$ & -1.26 &  -- & $24.6 \pm 0.58$ & $1.51 \pm 0.15$ & 10--1000 & $1990^{+47.3}_{-47.3}$ & $327^{+31.9}_{-31.9}$ & -- & (12) \\
		131108A & 2.4 & $18.18 \pm 0.57$ & -- & SBPL & $318.13^{+40.22}_{-40.22}$ & -1.04 & -2.42 & $36.2 \pm 0.42$ & $6.81 \pm 0.29$ & 10--1000 & $6330^{+73.1}_{-73.1}$ & $4040^{+175}_{-175}$ & -- & (12) \\
		131231A & 0.642 & $31.23 \pm 0.57$ & -- & Band & $178.09^{+4.03}_{-4.03}$ & -1.22 & -2.3 & $153.8 \pm 0.67$ & $9.51 \pm 0.31$ & 10--1000 & $2230^{+9.73}_{-9.73}$ & $226^{+7.33}_{-7.33}$ & -- & (12) \\
		140206A & 2.73 & $146.69 \pm 4.42$ & -- & SBPL & $537.61^{+153.06}_{-153.06}$ & -1.45 & -2.05 & $127.71 \pm 0.89$ & $6.41 \pm 0.22$ & 10--1000 & $30900^{+215}_{-215}$ & $5790^{+195}_{-195}$ & -- & (12) \\
		140213A & 1.2076 & $18.62 \pm 0.72$ & -- & Band & $86.15^{+4.1}_{-4.1}$ & -1.13 & -2.25 & $23.9 \pm 0.4$ & $3.92 \pm 0.19$ & 10--1000 & $1210^{+20.1}_{-20.1}$ & $437^{+21.2}_{-21.2}$ & -- & (12) \\
		140304A & 5.283 & $31.23 \pm 8.72$ & -- & CPL & $141.44^{+18.46}_{-18.46}$ & -0.89 &  -- & $1.92 \pm 0.16$ & $0.42 \pm 0.08$ & 10--1000 & $999^{+84.3}_{-84.3}$ & $1380^{+266}_{-266}$ & -- & (12) \\
		140423A & 3.26 & $95.23 \pm 11.59$ & -- & Band & $116.17^{+15.89}_{-15.89}$ & -0.55 & -1.79 & $21.5 \pm 0.6$ & $0.79 \pm 0.15$ & 10--1000 & $6880^{+193}_{-193}$ & $1070^{+211}_{-211}$ & -- & (12) \\
		140506A & 0.889 & $64.13 \pm 2.01$ & -- & CPL & $197.58^{+26.29}_{-26.29}$ & -1.18 &  -- & $4.63 \pm 0.33$ & $2.78 \pm 0.33$ & 10--1000 & $105^{+7.48}_{-7.48}$ & $119^{+14.1}_{-14.1}$ & -- & (12) \\
		140508A & 1.027 & $44.29 \pm 0.23$ & -- & Band & $257.4^{+12.12}_{-12.12}$ & -1.18 & -2.32 & $63.1 \pm 0.8$ & $11.95 \pm 0.59$ & 10--1000 & $2370^{+30.1}_{-30.1}$ & $910^{+44.6}_{-44.6}$ & -- & (12) \\
		140512A & 0.725 & $147.97 \pm 2.36$ & -- & CPL & $691.86^{+57.05}_{-57.05}$ & -1.23 &  -- & $45.5 \pm 0.73$ & $2.81 \pm 0.23$ & 10--1000 & $852^{+13.7}_{-13.7}$ & $90.6^{+7.41}_{-7.41}$ & -- & (12) \\
		140606B & 0.384 & $22.78 \pm 2.06$ & -- & CPL & $577.63^{+102.84}_{-102.84}$ & -1.24 &  -- & $9.57 \pm 0.43$ & $2.79 \pm 0.23$ & 10--1000 & $45^{+2}_{-2}$ & $18.2^{+1.51}_{-1.51}$ & -- & (12) \\
		140620A & 2.04 & $45.83 \pm 12.13$ & -- & CPL & $130.24^{+12.24}_{-12.24}$ & -1.29 &  -- & $6.17 \pm 0.29$ & $0.79 \pm 0.15$ & 10--1000 & $725^{+33.5}_{-33.5}$ & $281^{+52}_{-52}$ & -- & (12) \\
		140623A & 1.92 & $111.1 \pm 4$ & -- & CPL & $317.03^{+82.62}_{-82.62}$ & -1.42 &  -- & $3.36 \pm 0.35$ & $0.47 \pm 0.14$ & 10--1000 & $370^{+38.4}_{-38.4}$ & $151^{+45.3}_{-45.3}$ & -- & (12) \\
		140703A & 3.14 & $83.97 \pm 3$ & -- & CPL & $218.59^{+23.22}_{-23.22}$ & -1.28 &  -- & $9.04 \pm 0.42$ & $0.63 \pm 0.11$ & 10--1000 & $2180^{+100}_{-100}$ & $628^{+110}_{-110}$ & -- & (12) \\
		140801A & 1.32 & $7.17 \pm 0.57$ & -- & CPL & $121.11^{+2.11}_{-2.11}$ & -0.4 &  -- & $12 \pm 0.16$ & $3.84 \pm 0.18$ & 10--1000 & $560^{+7.24}_{-7.24}$ & $416^{+19.1}_{-19.1}$ & -- & (12) \\
		140808A & 3.29 & $4.48 \pm 0.36$ & -- & CPL & $125.43^{+6.15}_{-6.15}$ & -0.49 &  -- & $3.22 \pm 0.11$ & $1.18 \pm 0.14$ & 10--1000 & $773^{+27.1}_{-27.1}$ & $1210^{+148}_{-148}$ & -- & (12) \\
		140907A & 1.21 & $35.84 \pm 5.47$ & -- & CPL & $137.2^{+7.81}_{-7.81}$ & -1.01 &  -- & $6.23 \pm 0.22$ & $0.58 \pm 0.13$ & 10--1000 & $257^{+9.08}_{-9.08}$ & $53^{+11.9}_{-11.9}$ & -- & (12) \\
		141004A & 0.573 & $2.56 \pm 0.61$ & $276.05 \pm 48.98$ & CPL & $181.77^{+55.61}_{-55.61}$ & -1.6 &  -- & $1.42 \pm 0.18$ & $1.57 \pm 0.17$ & 10--1000 & $15.1^{+1.91}_{-1.91}$ & $26.2^{+2.85}_{-2.85}$ & -- & (12),(8) \\
		141028A & 2.33 & $31.49 \pm 2.43$ & -- & Band & $293.1^{+17.98}_{-17.98}$ & -0.84 & -1.97 & $39.9 \pm 0.4$ & $4.09 \pm 0.26$ & 10--1000 & $7470^{+75.5}_{-75.5}$ & $2550^{+164}_{-164}$ & -- & (12) \\
		141220A & 1.3195 & $7.62 \pm 0.92$ & -- & CPL & $178.3^{+9.11}_{-9.11}$ & -0.82 &  -- & $5.55 \pm 0.18$ & $1.96 \pm 0.18$ & 10--1000 & $262^{+8.36}_{-8.36}$ & $215^{+19.8}_{-19.8}$ & -- & (12) \\
		141221A & 1.452 & $23.81 \pm 1.72$ & -- & CPL & $182.07^{+31.97}_{-31.97}$ & -1.18 &  -- & $2.86 \pm 0.26$ & $0.96 \pm 0.16$ & 10--1000 & $171^{+15.7}_{-15.7}$ & $140^{+23.9}_{-23.9}$ & -- & (12) \\
		141225A & 0.915 & $56.32 \pm 4.89$ & -- & CPL & $257.58^{+26.96}_{-26.96}$ & -0.6 &  -- & $5.91 \pm 0.37$ & $0.78 \pm 0.2$ & 10--1000 & $134^{+8.32}_{-8.32}$ & $33.7^{+8.59}_{-8.59}$ & -- & (12) \\
		150101B & 0.134 & $0.08 \pm 0.93$ & $20.91 \pm 3.4$ & CPL & $125.11^{+48.57}_{-48.57}$ & -1.36 &  -- & $0.08 \pm 0.02$ & $0.87 \pm 0.11$ & 10--1000 & $0.04^{+0.01}_{-0.01}$ & $0.47^{+0.06}_{-0.06}$ & -- & (12),(8) \\
		150301B & 1.5169 & $13.31 \pm 1.56$ & -- & CPL & $225.43^{+27.53}_{-27.53}$ & -1.12 &  -- & $4.04 \pm 0.25$ & $0.78 \pm 0.13$ & 10--1000 & $258^{+15.7}_{-15.7}$ & $125^{+20.6}_{-20.6}$ & -- & (12) \\
		150314A & 1.758 & $10.69 \pm 0.14$ & -- & Band & $347.16^{+7.9}_{-7.9}$ & -0.68 & -2.6 & $88.9 \pm 0.61$ & $16.3 \pm 0.44$ & 10--1000 & $8670^{+59.6}_{-59.6}$ & $4380^{+118}_{-118}$ & -- & (12) \\
		150403A & 2.06 & $22.27 \pm 0.81$ & -- & Band & $428.74^{+21.06}_{-21.06}$ & -0.87 & -2.11 & $64.2 \pm 0.64$ & $8.2 \pm 0.45$ & 10--1000 & $9700^{+96.1}_{-96.1}$ & $3790^{+208}_{-208}$ & -- & (12) \\
		150514A & 0.807 & $10.81 \pm 1.07$ & -- & CPL & $78.45^{+4.22}_{-4.22}$ & -1.36 &  -- & $4.57 \pm 0.13$ & $1.52 \pm 0.1$ & 10--1000 & $96.4^{+2.72}_{-2.72}$ & $58^{+3.93}_{-3.93}$ & -- & (12) \\
		150727A & 0.313 & $49.41 \pm 3.97$ & -- & CPL & $208.53^{+17.14}_{-17.14}$ & -0.35 &  -- & $5.9 \pm 0.32$ & $0.77 \pm 0.2$ & 10--1000 & $14.3^{+0.78}_{-0.78}$ & $2.43^{+0.64}_{-0.64}$ & -- & (12) \\
		150818A & 0.282 & $184.26 \pm 18.43$ & $568.98 \pm 99.37$ & CPL & $111^{+53}_{-26}$ & -1.77 &  -- & $8.1 \pm 1.09$ & $0.55 \pm 0.15$ & 15--15000 & $23^{+3.09}_{-3.09}$ & $2^{+0.54}_{-0.54}$ & SN & (16),(8) \\
		150821A & 0.755 & $103.43 \pm 5.75$ & -- & Band & $281.22^{+17.11}_{-17.11}$ & -1.24 & -2.13 & $71.1 \pm 0.83$ & $1.86 \pm 0.19$ & 10--1000 & $1600^{+18.6}_{-18.6}$ & $73.3^{+7.6}_{-7.6}$ & -- & (12) \\
		151027A & 0.81 & $123.39 \pm 1.15$ & -- & SBPL & $304.98^{+267.4}_{-267.4}$ & -1.35 & -2.04 & $17.3 \pm 0.45$ & $1.67 \pm 0.18$ & 10--1000 & $466^{+12.2}_{-12.2}$ & $81.5^{+8.53}_{-8.53}$ & -- & (12) \\
		160509A & 1.17 & $369.67 \pm 0.81$ & -- & Band & $355.19^{+9.88}_{-9.88}$ & -1.02 & -2.23 & $204.14 \pm 1.11$ & $14.8 \pm 0.41$ & 10--1000 & $10500^{+57.1}_{-57.1}$ & $1650^{+45.8}_{-45.8}$ & -- & (12) \\
		160625B & 1.406 & $453.39 \pm 0.57$ & -- & SBPL & $510.94^{+27.12}_{-27.12}$ & -1.02 & -2.1 & $668.22 \pm 1.99$ & $48.14 \pm 0.86$ & 10--1000 & $53200^{+158}_{-158}$ & $9220^{+165}_{-165}$ & -- & (12) \\
		160629A & 3.332 & $64.77 \pm 0.92$ & -- & CPL & $290.75^{+19.27}_{-19.27}$ & -1.03 &  -- & $18.9 \pm 0.61$ & $1.13 \pm 0.15$ & 10--1000 & $4830^{+156}_{-156}$ & $1250^{+166}_{-166}$ & -- & (12) \\
		160804A & 0.736 & $131.59 \pm 21.72$ & -- & CPL & $75.85^{+2.81}_{-2.81}$ & -1.09 &  -- & $12.4 \pm 0.29$ & $0.48 \pm 0.11$ & 10--1000 & $203^{+4.65}_{-4.65}$ & $13.7^{+3.11}_{-3.11}$ & -- & (12) \\
		160821B & 0.16 & $1.09 \pm 0.98$ & $26.53 \pm 0.71$ & CPL & $91.97^{+27.87}_{-27.87}$ & -1.4 &  -- & $0.17 \pm 0.02$ & $0.65 \pm 0.08$ & 10--1000 & $0.12^{+0.02}_{-0.02}$ & $0.54^{+0.07}_{-0.07}$ & -- & (12),(8) \\
		161001A & 0.67 & $2.24 \pm 0.23$ & $32.57 \pm 5.93$ & CPL & $373.17^{+58.58}_{-58.58}$ & -0.94 &  -- & $2.36 \pm 0.19$ & $3.27 \pm 0.23$ & 10--1000 & $30.2^{+2.4}_{-2.4}$ & $69.8^{+4.97}_{-4.97}$ & -- & (12),This work \\
		161014A & 2.823 & $36.61 \pm 1.49$ & -- & CPL & $169.72^{+12.78}_{-12.78}$ & -0.76 &  -- & $5.08 \pm 0.26$ & $1.14 \pm 0.16$ & 10--1000 & $950^{+48.8}_{-48.8}$ & $812^{+111}_{-111}$ & -- & (12) \\
		161017A & 2.013 & $37.89 \pm 10.86$ & -- & CPL & $277.18^{+40.6}_{-40.6}$ & -1.09 &  -- & $5.79 \pm 0.4$ & $1.4 \pm 0.3$ & 10--1000 & $626^{+42.7}_{-42.7}$ & $455^{+97}_{-97}$ & -- & (12) \\
		161117A & 1.549 & $122.18 \pm 0.66$ & -- & CPL & $85.04^{+1.72}_{-1.72}$ & -0.87 &  -- & $30.8 \pm 0.4$ & $1.08 \pm 0.17$ & 10--1000 & $2080^{+26.9}_{-26.9}$ & $186^{+29.2}_{-29.2}$ & -- & (12) \\
		161129A & 0.645 & $36.1 \pm 0.72$ & -- & CPL & $214.04^{+22.38}_{-22.38}$ & -1.16 &  -- & $8.85 \pm 0.44$ & $1.41 \pm 0.17$ & 10--1000 & $104^{+5.12}_{-5.12}$ & $27.1^{+3.29}_{-3.29}$ & -- & (12) \\
		161219B & 0.1475 & $7.04 \pm 0.7$ & -- & CPL & $94^{+19}_{-12}$ & -1.51 &  -- & $2.69 \pm 0.3$ & $0.88 \pm 0.16$ & 15--15000 & $1.81^{+0.2}_{-0.2}$ & $0.68^{+0.12}_{-0.12}$ & SN & (16) \\
		170113A & 1.968 & $49.15 \pm 4.14$ & -- & CPL & $106.56^{+36.71}_{-36.71}$ & -1.69 &  -- & $1.74 \pm 0.3$ & $0.5 \pm 0.17$ & 10--1000 & $234^{+39.7}_{-39.7}$ & $198^{+67.7}_{-67.7}$ & -- & (12) \\
		170214A & 2.53 & $122.88 \pm 0.72$ & -- & SBPL & $402.03^{+27.05}_{-27.05}$ & -1.06 & -2.25 & $197.64 \pm 1.08$ & $4.75 \pm 0.37$ & 10--1000 & $40000^{+218}_{-218}$ & $3390^{+262}_{-262}$ & -- & (12) \\
		170607A & 0.557 & $20.93 \pm 2.1$ & -- & CPL & $145.17^{+11.88}_{-11.88}$ & -1.4 &  -- & $9.44 \pm 0.32$ & $1.73 \pm 0.22$ & 10--1000 & $87.6^{+3}_{-3}$ & $25^{+3.24}_{-3.24}$ & -- & (12) \\
		170705A & 2.01 & $22.78 \pm 1.38$ & -- & Band & $97.88^{+7.64}_{-7.64}$ & -0.99 & -2.3 & $14.4 \pm 0.43$ & $2.64 \pm 0.24$ & 10--1000 & $1780^{+53.5}_{-53.5}$ & $986^{+90.2}_{-90.2}$ & -- & (12) \\
		170817A & 0.0093 & $2.05 \pm 0.47$ & $150 \pm 15$ & CPL & $215.09^{+54.22}_{-54.22}$ & 0.14 &  -- & $0.14 \pm 0.03$ & $0.73 \pm 0.18$ & 10--1000 & $0.0003^{+0.0001}_{-0.0001}$ & $0.0014^{+0.0003}_{-0.0003}$ & -- & (12) \\
		171010A & 0.3285 & $107.27 \pm 0.81$ & -- & Band & $137.66^{+1.43}_{-1.43}$ & -1.09 & -2.19 & $671.82 \pm 1.66$ & $15.7 \pm 0.5$ & 10--1000 & $2530^{+6.26}_{-6.26}$ & $78.6^{+2.53}_{-2.53}$ & -- & (12) \\
		171205A & 0.0368 & $216.9 \pm 21.69$ & $9560 \pm 3130$ & CPL & $111^{+23}_{-15}$ & -0.78 &  -- & $7.43 \pm 1.13$ & $0.57 \pm 0.15$ & 15--15000 & $0.24^{+0.04}_{-0.04}$ & $0.02^{+0.01}_{-0.01}$ & SN & (16),(8) \\
		171222A & 2.409 & $80.38 \pm 4.62$ & -- & CPL & $203.5^{+3.44}_{-3.44}$ & -2.01 &  -- & $2.02 \pm 0.1$ & $0.42 \pm 0.14$ & 10--1000 & $564^{+27}_{-27}$ & $402^{+132}_{-132}$ & -- & (12) \\
		180618A & 0.544 & $3.71 \pm 0.58$ & $3.26 \pm 10.22$ & CPL & $2507.58^{+917.74}_{-917.74}$ & -1.13 &  -- & $2.15 \pm 0.14$ & $3.78 \pm 0.48$ & 10--1000 & $42.6^{+2.82}_{-2.82}$ & $116^{+14.8}_{-14.8}$ & -- & (12),(8) \\
		180620B & 1.1175 & $46.72 \pm 1.33$ & -- & Band & $175.63^{+49.79}_{-49.79}$ & -1.21 & -1.66 & $12.5 \pm 0.34$ & $2.03 \pm 0.44$ & 10--1000 & $817^{+22.5}_{-22.5}$ & $281^{+60.5}_{-60.5}$ & -- & (12) \\
		180720B & 0.654 & $48.9 \pm 0.36$ & -- & Band & $636.04^{+15.43}_{-15.43}$ & -1.17 & -2.49 & $317.71 \pm 1.45$ & $32.45 \pm 0.51$ & 10--1000 & $5310^{+24.2}_{-24.2}$ & $897^{+14}_{-14}$ & -- & (12) \\
		180728A & 0.117 & $6.4 \pm 0.36$ & $-71.2 \pm 32.08$ & Band & $79.2^{+1.4}_{-1.4}$ & -1.54 & -2.46 & $56.5 \pm 0.2$ & $19.24 \pm 0.36$ & 10--1000 & $24.6^{+0.09}_{-0.09}$ & $9.36^{+0.17}_{-0.17}$ & SN & (12),(8) \\
		181201A & 0.45 & $172 \pm 17.2$ & -- & Band & $152^{+6}_{-6}$ & -1.25 & -2.73 & $199 \pm 6$ & $28.8 \pm 3.4$ & 20--10000 & $1180^{+35.7}_{-35.7}$ & $248^{+29.3}_{-29.3}$ & SN & (17) \\
		190114C & 0.425 & $116.35 \pm 2.56$ & $745.25 \pm 86.39$ & Band & $998.6^{+11.9}_{-11.9}$ & -1.06 & -3.18 & $399 \pm 0.81$ & $84.19 \pm 0.97$ & 10--1000 & $3010^{+6.11}_{-6.11}$ & $904^{+10.4}_{-10.4}$ & -- & (12),(8) \\
		200826A & 0.7481 & $1.14 \pm 0.13$ & $57.87 \pm 0.5$ & Band & $89.8^{+3.7}_{-3.7}$ & -0.41 & -2.4 & $4.8 \pm 0.1$ & $7 \pm 0.22$ & 10--1000 & $85^{+1.77}_{-1.77}$ & $217^{+6.92}_{-6.92}$ & -- & (12),This work \\
		211023A & 0.39 & $79.11 \pm 0.57$ & -- & Band & $92^{+2}_{-2}$ & -1.74 & -2.55 & $111 \pm 0.7$ & $2.91 \pm 5.04$ & 10--1000 & $629^{+3.97}_{-3.97}$ & $23^{+39.7}_{-39.7}$ & -- & (12) \\
		211211A & 0.076 & $34.31 \pm 0.57$ & $12 \pm 10$ & Band & $646.8^{+7.8}_{-7.8}$ & -1.3 & -2.4 & $540 \pm 1$ & $79.19 \pm 1.19$ & 10--1000 & $111^{+0.21}_{-0.21}$ & $17.6^{+0.26}_{-0.26}$ & -- & (12),(18) \\
		211211A$^{*}$ & 0.076 & $13 \pm 1.3$ & $10 \pm 3$ & Band & $687.1^{+12.55}_{-11}$ & -1 & -2.36 & $377 \pm 1$ & $111.32 \pm 1.67$ & 10--1000 & $83^{+0.22}_{-0.22}$ & $26.4^{+0.4}_{-0.4}$ & -- & (18) \\
		211227A & 0.228 & $82.5 \pm 8.25$ & -- & Band & $192^{+45}_{-42}$ & -1.34 & -2.26 & $26 \pm 2.1$ & $2 \pm 0.4$ & 15--1500 & $44.2^{+3.57}_{-3.57}$ & $4.18^{+0.84}_{-0.84}$ & KN & (19) \\
		211227A$^{*}$ & 0.228 & $4 \pm 0.4$ & -- & CPL & $400^{+1200}_{-200}$ & -1.56 &  -- & $2.01 \pm 0.19$ & $2 \pm 0.4$ & 15--1500 & $3.11^{+0.29}_{-0.29}$ & $3.8^{+0.76}_{-0.76}$ & KN & (20) \\
		230307A & 0.065 & $34.56 \pm 0.57$ & $12.94 \pm 0.01$ & CPL & $936^{+3}_{-3}$ & -1.07 &  -- & $2950 \pm 4$ & $226.42 \pm 3.93$ & 10--1000 & $427^{+0.58}_{-0.58}$ & $34.9^{+0.61}_{-0.61}$ & -- & (12),This work \\
		230812B & 0.36 & $3.26 \pm 0.09$ & -- & Band & $273^{+3}_{-3}$ & -0.8 & -2.47 & $252 \pm 0.02$ & $161.72 \pm 1.52$ & 10--1000 & $1070^{+0.08}_{-0.08}$ & $932^{+8.78}_{-8.78}$ & -- & (21) \\
		\enddata
		
		\tablerefs{
			(1)-\cite{2017ApJ...850..161T};(2)-\cite{2002AA...390...81A};(3)-\cite{2003ApJ...594L..79Y};(4)-\cite{2013ApJS..208...21G};(5)-\cite{2004ApJ...602..875S};(6)-\cite{2008AA...481..319W};(7)-\cite{2016ApJ...829....7L};(8)-\cite{2023MNRAS.524.1096L};(9)-\cite{2007ApJ...654..385K};(10)-\cite{2023ApJ...950...30Z};(11)-\cite{2006Natur.444.1044G};(12)-\cite{2020ApJ...893...46V};(13)-\cite{2008GCN..8369....1B};(14)-\cite{2011GCN.12663....1G};(15)-\cite{2013GCN.15316....1S};(16)-\cite{2021ApJ...908...83T};(17)-\cite{2018GCN.23495....1S};(18)-\cite{2022Natur.612..232Y};(19)-\cite{2022GCN.31544....1T};(20)-\cite{2022ApJ...936L..10Z};(21)-\cite{2023GCN.34391....1R}.}
		
	\end{deluxetable*}
\end{longrotatetable}

\begin{longrotatetable}
	\begin{deluxetable*}{lcccc ccccc ccccc cc}
		\tablecaption{The prompt emission parameters of peculiar GRBs}
		\label{table:paticular}
		\tabletypesize{\tiny}
		
		\tablehead{
			\colhead{$GRB$} & \colhead{$T_{90}$} & \colhead{$model$} & \colhead{$E_{\rm p}$} & \colhead{$\alpha$} & \colhead{$\beta$} & \colhead{$S_{\gamma,-6}$} & \colhead{$F_{\rm p}$} & \colhead{$HR$} & \colhead{$\rm res^{a}$} & \colhead{$\tau_{31}$} & \colhead{$P^{b}_{E_{\rm p}}$} & \colhead{$P^{b}_{\rm HR}$} & \colhead{$z^{c}_{\varepsilon}$} & \colhead{$P^{d}_{\varepsilon}$} & \colhead{$z^{c}_{\rm EH}$} & \colhead{$P^{d}_{\rm EH}$}
			\\
			\colhead{} & \colhead{(s)} & \colhead{}  & \colhead{(keV)}  & \colhead{}  & \colhead{}  & \colhead{(erg cm$^{-2}$)} & \colhead{(ph cm$^{-2}$ s$^{-1}$)}  & \colhead{}  & \colhead{(ms)}  & \colhead{(ms)}  & \colhead{(\%)} & \colhead{(\%)}) & \colhead{} & \colhead{(\%)} &  & \colhead{(\%)}\\
		}
		\startdata
		Group A GRBs\\
		\hline
		080816989 & $4.61 \pm 0.45$ & CPL & $1530 \pm 224$ & -0.5 &\nodata& $3.21 \pm 0.15$ & $9.27 \pm 0.62$ & 2.23 & 64 & $-110.17 \pm 19.11$ & 8.9 & 35.3 &\nodata& 0 &\nodata& 0 \\
		080828189 & $3.01 \pm 3.33$ & CPL & $118.9 \pm 10.3$ & 10.65 &\nodata& $0.11 \pm 0.02$ & $5.62 \pm 0.98$ & 1.51 & 64 & $-134.49 \pm 40.63$ & 81 & 28.2 &\nodata& 0 &\nodata& 0 \\
		081006604 & $6.4 \pm 0.92$ & CPL & $653.6 \pm 165.8$ & -0.43 &\nodata& $1.72 \pm 0.18$ & $4.69 \pm 1.2$ & 1.23 & 256 &  \nodata & 62.3 & 83.1 &\nodata& 0 &\nodata& 0 \\
		090510325 & $7.42 \pm 1.72$ & CPL & $1957 \pm 581$ & -0.63 &\nodata& $1.84 \pm 0.17$ & $3.14 \pm 1.21$ & 1.18 & 256 &  \nodata & 18.6 & 88.6 &\nodata& 0 &\nodata& 0 \\
		090518080 & $2.05 \pm 0.41$ & CPL & $331.5 \pm 122.1$ & -1.35 &\nodata& $1 \pm 0.16$ & $9.75 \pm 5.05$ & 1.01 & 64 & $101.92 \pm 24.86$ & 17.9 & 19.3 & 2.727--10 & 20.4 &\nodata& 0 \\
		090610648 & $6.14 \pm 8.14$ & CPL & $2000 \pm 416$ & -0.79 &\nodata& $1.89 \pm 0.11$ & $4.6 \pm 0.74$ & 2.81 & 128 & $-467.66 \pm 179.86$ & 11 & 38.8 &\nodata& 0 &\nodata& 0 \\
		100302061 & $2.24 \pm 1.61$ & CPL & $2739 \pm 664$ & -0.75 &\nodata& $0.69 \pm 0.06$ & $4.28 \pm 1.17$ & 0.81 & 128 & $-174.89 \pm 119.58$ & 0.2 & 28.1 &\nodata& 0 &\nodata& 0 \\
		100516369 & $2.11 \pm 1.13$ & CPL & $208.8 \pm 57.4$ & -0.36 &\nodata& $0.07 \pm 0.02$ & $5.09 \pm 1.06$ & 1.41 & 64 & $313.66 \pm 114.63$ & 36.5 & 13.3 &\nodata& 0 &\nodata& 0 \\
		100717372 & $5.95 \pm 1.51$ & CPL & $3994 \pm 1288$ & -1.01 &\nodata& $1.29 \pm 0.09$ & $7.57 \pm 1.36$ & 1.24 & 16 & $18.07 \pm 5.35$ & 2 & 79.9 &\nodata& 0 &\nodata& 0 \\
		100717446 & $2.43 \pm 1.36$ & CPL & $692.8 \pm 192.3$ & -0.62 &\nodata& $0.74 \pm 0.06$ & $5.56 \pm 1.03$ & 1.95 & 128 & $358.59 \pm 53.72$ & 7.5 & 10 &\nodata& 0 &\nodata& 0 \\
		100719825 & $3.07 \pm 3.11$ & CPL & $405.5 \pm 102.5$ & -0.55 &\nodata& $1.32 \pm 0.2$ & $4.57 \pm 1.45$ & 0.93 & 128 & $203.01 \pm 119.59$ & 33.5 & 46.7 & 4--8.455 & 9.5 &\nodata& 0 \\
		100922625 & $4.35 \pm 0.92$ & CPL & $1338 \pm 343$ & -0.56 &\nodata& $1.19 \pm 0.13$ & $2.55 \pm 1.03$ & 0.8 & 256 & $-163.48 \pm 92.23$ & 10.1 & 72.5 &\nodata& 0 &\nodata& 0 \\
		101214748 & $2.24 \pm 2.08$ & CPL & $345.3 \pm 100.5$ & -0.52 &\nodata& $0.23 \pm 0.05$ & $7.41 \pm 1.38$ & 0.76 & 32 & $80.82 \pm 33.56$ & 21 & 29.1 &\nodata& 0 &\nodata& 0 \\
		110112934 & $2.3 \pm 2.54$ & CPL & $552.1 \pm 138.3$ & -0.74 &\nodata& $0.4 \pm 0.06$ & $8.22 \pm 1.23$ & 0.9 & 32 & $35.58 \pm 157.44$ & 10 & 28 &\nodata& 0 &\nodata& 0 \\
		110307972 & $2.3 \pm 3.44$ & CPL & $347.6 \pm 57.9$ & 0.42 &\nodata& $0.49 \pm 0.07$ & $5.88 \pm 0.98$ & 1.91 & 256 & $-978.2 \pm 55.71$ & 22.2 & 8.9 &\nodata& 0 &\nodata& 0 \\
		110331604 & $3.2 \pm 0.95$ & CPL & $905 \pm 201.1$ & -0.49 &\nodata& $2.24 \pm 0.15$ & $4.01 \pm 1.3$ & 1.71 & 256 & $1041 \pm 211.1$ & 9.6 & 26.5 &\nodata& 0 &\nodata& 0 \\
		110401920 & $2.37 \pm 1.27$ & CPL & $1282 \pm 366$ & -0.76 &\nodata& $1.37 \pm 0.09$ & $10.81 \pm 1.37$ & 2.73 & 16 & $-65.3 \pm 9.17$ & 1.8 & 3.5 &\nodata& 0 &\nodata& 0 \\
		120114433 & $2.75 \pm 1.57$ & CPL & $240.8 \pm 62.2$ & 0.38 &\nodata& $0.19 \pm 0.06$ & $3.78 \pm 1.06$ & 1.1 & 256 &  \nodata & 49.5 & 34 &\nodata& 0 &\nodata& 0 \\
		120219563 & $8.13 \pm 0.43$ & CPL & $1990 \pm 750$ & -0.82 &\nodata& $1.44 \pm 0.13$ & $2.6 \pm 0.83$ & 1.55 & 256 &  \nodata & 22.2 & 86.4 &\nodata& 0 &\nodata& 0 \\
		121220311 & $5.12 \pm 0.81$ & CPL & $739.5 \pm 266.8$ & -0.67 &\nodata& $0.97 \pm 0.13$ & $2.73 \pm 0.93$ & 1.13 & 128 & $257.38 \pm 213.61$ & 40.2 & 74.9 &\nodata& 0 &\nodata& 0 \\
		130112353 & $2.05 \pm 1.56$ & CPL & $548.3 \pm 90.1$ & -0.71 &\nodata& $1.22 \pm 0.09$ & $8.6 \pm 1.17$ & 1.23 & 32 & $226.49 \pm 47.69$ & 7.2 & 15.1 &\nodata& 0 &\nodata& 0 \\
		130518551 & $4.1 \pm 2.57$ & CPL & $980.5 \pm 173.5$ & -0.83 &\nodata& $4.03 \pm 0.18$ & $19.81 \pm 1.08$ & 1.45 & 8 & $-0.02 \pm 0.82$ & 15.9 & 51.4 &\nodata& 0 &\nodata& 0 \\
		131123543 & $3.14 \pm 0.72$ & CPL & $374.1 \pm 82.9$ & -0.77 &\nodata& $1 \pm 0.16$ & $6.65 \pm 1.16$ & 1.08 & 64 & $-320.91 \pm 290.05$ & 38.6 & 43.9 &\nodata& 0 &\nodata& 0 \\
		140216331 & $2.43 \pm 0.59$ & CPL & $779.2 \pm 284.4$ & -0.84 &\nodata& $1.08 \pm 0.13$ & $3.46 \pm 1.04$ & 1.98 & 128 &  \nodata & 5.9 & 9.6 &\nodata& 0 &\nodata& 0 \\
		140224382 & $2.3 \pm 2.11$ & CPL & $1329 \pm 479$ & -0.99 &\nodata& $2 \pm 0.16$ & $8.81 \pm 1.35$ & 0.96 & 16 & $-13.52 \pm 17.94$ & 1.5 & 26.7 &\nodata& 0 &\nodata& 0 \\
		140619475 & $2.82 \pm 0.81$ & CPL & $1479 \pm 261$ & -0.26 &\nodata& $1.67 \pm 0.09$ & $5.28 \pm 0.75$ & 3.21 & 128 & $208.47 \pm 76.18$ & 2.2 & 3.5 &\nodata& 0 &\nodata& 0 \\
		140831374 & $3.58 \pm 2.57$ & CPL & $836.8 \pm 313.9$ & -1.06 &\nodata& $1.58 \pm 0.16$ & $6.08 \pm 1.4$ & 1.15 & 64 & $21.97 \pm 73.44$ & 15.1 & 51.5 &\nodata& 0 &\nodata& 0 \\
		140930134 & $3.26 \pm 1.62$ & CPL & $797.5 \pm 317.9$ & -0.69 &\nodata& $1.03 \pm 0.11$ & $4.73 \pm 0.89$ & 1.7 & 256 &  \nodata & 12.9 & 27.8 &\nodata& 0 &\nodata& 0 \\
		141102536 & $2.62 \pm 0.33$ & CPL & $639.4 \pm 99.1$ & -0.7 &\nodata& $1.61 \pm 0.08$ & $15.59 \pm 1.24$ & 1.34 & 8 & $-9.94 \pm 1.51$ & 10.9 & 24.6 &\nodata& 0 &\nodata& 0 \\
		141222298 & $2.75 \pm 0.26$ & CPL & $3290 \pm 956$ & -1.55 &\nodata& $9.25 \pm 0.16$ & $121.02 \pm 2.62$ & 0.45 & 8 & $1.56 \pm 0.04$ & 0.3 & 43.2 &\nodata& 0 &\nodata& 0 \\
		150228845 & $4.13 \pm 1.06$ & CPL & $1184 \pm 146.7$ & -0.82 &\nodata& $7.48 \pm 0.17$ & $17.76 \pm 1.32$ & 1.6 & 16 & $80.15 \pm 27.26$ & 11.2 & 47.1 & 1.727--8.182 & 35.9 &\nodata& 0 \\
		150828901 & $2.05 \pm 1.78$ & CPL & $589.9 \pm 209$ & -0.95 &\nodata& $0.56 \pm 0.11$ & $3.01 \pm 0.9$ & 2.59 & 256 &  \nodata & 6.3 & 2.7 &\nodata& 0 &\nodata& 0 \\
		160219289 & $3.52 \pm 0.18$ & CPL & $547.6 \pm 133$ & -0.87 &\nodata& $1.48 \pm 0.16$ & $19.19 \pm 1.11$ & 1.18 & 16 & $77.07 \pm 30.88$ & 29.3 & 49 &\nodata& 0 &\nodata& 0 \\
		160628136 & $5.89 \pm 2.43$ & CPL & $943 \pm 307$ & -0.94 &\nodata& $2.01 \pm 0.16$ & $6.13 \pm 1.29$ & 1.35 & 128 &  \nodata & 37.6 & 77.1 &\nodata& 0 &\nodata& 0 \\
		160818230 & $2.18 \pm 0.91$ & CPL & $256.7 \pm 50.8$ & -0.4 &\nodata& $0.44 \pm 0.06$ & $8.18 \pm 2.05$ & 0.96 & 32 & $-100.95 \pm 18.01$ & 30 & 23.3 &\nodata& 0 &\nodata& 0 \\
		161001045 & $2.24 \pm 0.23$ & CPL & $373.2 \pm 58.6$ & -0.94 &\nodata& $2.36 \pm 0.19$ & $17.61 \pm 1.25$ & 1.01 & 8 & $32.57 \pm 5.93$ & 18.5 & 23.8 & 1.273--10 & 49.6 &\nodata& 0 \\
		161210524 & $2.3 \pm 2.86$ & CPL & $361.5 \pm 106.8$ & -0.94 &\nodata& $1.08 \pm 0.2$ & $3.9 \pm 0.91$ & 0.91 & 64 & $24.67 \pm 44.67$ & 20.9 & 27.7 & 4--8.727 & 9.7 &\nodata& 0 \\
		170121067 & $2.3 \pm 0.29$ & CPL & $1626 \pm 281$ & -0.75 &\nodata& $3.68 \pm 0.17$ & $10.29 \pm 0.87$ & 2.22 & 8 & $-6.03 \pm 2.1$ & 0.9 & 6.1 &\nodata& 0 &\nodata& 0 \\
		170802638 & $2.24 \pm 0.14$ & CPL & $724.1 \pm 148.7$ & -0.81 &\nodata& $2.6 \pm 0.13$ & $37.9 \pm 1.6$ & 1.69 & 8 & $23.01 \pm 10.86$ & 5.4 & 11 &\nodata& 0 &\nodata& 0 \\
		170817908 & $2.62 \pm 0.18$ & CPL & $1202 \pm 217$ & -0.93 &\nodata& $5.44 \pm 0.18$ & $10.37 \pm 1.61$ & 1.3 & 8 & $-21.82 \pm 2.12$ & 2.9 & 25.7 &\nodata& 0 &\nodata& 0 \\
		180618030 & $3.71 \pm 0.58$ & CPL & $2508 \pm 918$ & -1.13 &\nodata& $2.15 \pm 0.14$ & $14.61 \pm 1.86$ & 1.57 & 8 & $3.26 \pm 10.22$ & 1.5 & 40.5 &\nodata& 0 &\nodata& 0 \\
		\hline
		Group B GRBs\\
		\hline
		081107321 & $1.66 \pm 0.23$ & CPL & $70.8 \pm 2.3$ & -0.38 &\nodata& $1.25 \pm 0.04$ & $13.54 \pm 0.85$ & 0.47 & 64 & $82.84 \pm 2.44$ & 62.9 & 15.6 & 0.291--10 & 92.7 & 0.099--9.909 & 98 \\
		100816026 & $2.05 \pm 0.23$ & CPL & $140.1 \pm 6.1$ & -0.37 &\nodata& $3.35 \pm 0.11$ & $19.88 \pm 1.08$ & 1.1 & 16 & $137.06 \pm 7.38$ & 51.2 & 17.5 & 0.318--10 & 91.6 & 0.136--7.636 & 96 \\
		101216721 & $1.92 \pm 0.55$ & CPL & $152.9 \pm 8.2$ & -0.86 &\nodata& $2.92 \pm 0.1$ & $25.08 \pm 1.84$ & 0.76 & 16 & $121.75 \pm 7.57$ & 42.7 & 20.6 & 0.373--10 & 89.1 & 0.182--6.091 & 92.8 \\
		120222021 & $1.09 \pm 0.14$ & CPL & $130 \pm 6.9$ & -0.65 &\nodata& $1.72 \pm 0.06$ & $24.02 \pm 1.71$ & 0.76 & 16 & $64.52 \pm 2.68$ & 14.6 & 4.3 & 0.436--10 & 86 & 0.327--3.545 & 78 \\
		120323507 & $0.38 \pm 0.04$ & SBPL & $42.2 \pm 3.8$ & -1.17 & -2.08 & $10.7 \pm 0.13$ & $603.19 \pm 8.22$ & 0.44 & 4 & $20.17 \pm 0.1$ & 1 & 0.2 & 0.046--10 & 98.4 & 0.034--10 & 98.4 \\
		140209313 & $1.41 \pm 0.26$ & Band & $143.9 \pm 7.5$ & -0.61 & -2.4 & $9.52 \pm 0.18$ & $120.53 \pm 3.44$ & 0.93 & 16 & $92.03 \pm 0.34$ & 24.3 & 7.7 & 0.164--10 & 97 & 0.086--9.909 & 98.1 \\
		140626843 & $1.79 \pm 1.06$ & CPL & $176.1 \pm 22.5$ & -0.8 &\nodata& $1.03 \pm 0.09$ & $10.06 \pm 2.12$ & 0.9 & 16 & $887.64 \pm 195.02$ & 32.1 & 15.4 & 0.891--10 & 64.2 & 0.673--1.818 & 39.3 \\
		150923864 & $1.79 \pm 0.09$ & CPL & $143.5 \pm 7.6$ & -0.24 &\nodata& $1.3 \pm 0.05$ & $16.51 \pm 1.08$ & 1.1 & 16 & $28.76 \pm 14.62$ & 40.4 & 12.3 & 0.591--10 & 78.3 & 0.309--3.727 & 80 \\
		160806584 & $1.66 \pm 0.45$ & CPL & $138.7 \pm 9.7$ & -0.56 &\nodata& $1.37 \pm 0.07$ & $18.53 \pm 2.1$ & 0.81 & 64 & $191.29 \pm 4.39$ & 36.5 & 13.8 & 0.545--10 & 80.6 & 0.291--3.909 & 81.8 \\
		170206453 & $1.17 \pm 0.1$ & Band & $348 \pm 15.2$ & -0.46 & -2.59 & $10.8 \pm 0.16$ & $56.95 \pm 2.07$ & 2.12 & 8 & $116.66 \pm 1.77$ & 3.2 & 0.8 & 0.336--10 & 90.8 & 0.427--2.545 & 63.9 \\
		171126235 & $1.47 \pm 0.14$ & SBPL & $59.9 \pm 7$ & -0.68 & -2.65 & $2.55 \pm 0.09$ & $97.21 \pm 6.07$ & 0.12 & 8 & $42.06 \pm 0.99$ & 51.1 & 1.1 & 0.164--10 & 97 & 0.055--10 & 98.3 \\
		180511437 & $1.98 \pm 0.97$ & CPL & $107.5 \pm 11.9$ & -0.92 &\nodata& $0.61 \pm 0.04$ & $7.64 \pm 1.2$ & 0.51 & 64 & $-19.16 \pm 44.21$ & 59.8 & 24 & 0.691--10 & 73.4 & 0.273--4.182 & 84 \\
		180703949 & $1.54 \pm 0.09$ & CPL & $136.9 \pm 3$ & -0.77 &\nodata& $8.6 \pm 0.12$ & $109.24 \pm 3.09$ & 0.8 & 8 & $62.05 \pm 0.49$ & 31.5 & 11.3 & 0.182--10 & 96.5 & 0.09--10 & 98.1 \\
		\hline
		Group C GRBs\\
		\hline
		081130212 & $2.24 \pm 1$ & CPL & $27.7 \pm 2.8$ & 0.45 &\nodata& $0.21 \pm 0.02$ & $11.28 \pm 1.75$ & 0.14 & 128 &  \nodata & 93.4 & 5.6 & 0.318--10 & 91.6 & 0.057--10 & 98.3 \\
		090320045 & $2.37 \pm 0.27$ & CPL & $304.9 \pm 118.2$ & -1.09 &\nodata& $0.45 \pm 0.11$ & $2.8 \pm 0.65$ & 0.92 & 128 &  \nodata & 28.6 & 29.2 &\nodata& 0 &\nodata& 0 \\
		101002279 & $7.17 \pm 2.29$ & CPL & $547.1 \pm 210.4$ & -1.04 &\nodata& $1.24 \pm 0.21$ & $2.87 \pm 0.93$ & 0.83 & 256 & $-420.61 \pm 183.76$ & 76.5 & 90.8 &\nodata& 0 &\nodata& 0 \\
		120504945 & $5.76 \pm 0.78$ & CPL & $425.2 \pm 79.1$ & -0.48 &\nodata& $1.51 \pm 0.19$ & $4.98 \pm 1.33$ & 2.09 & 256 & $760.86 \pm 204.07$ & 74.1 & 55.5 & 3.364--10 & 14.4 &\nodata& 0 \\
		131128629 & $1.98 \pm 0.54$ & CPL & $59.2 \pm 4.6$ & 0.26 &\nodata& $0.22 \pm 0.02$ & $5.64 \pm 1.23$ & 0.39 & 256 & $164.35 \pm 63$ & 78.7 & 21.8 & 0.736--10 & 71.2 & 0.2--5.455 & 90.9 \\
		140912664 & $2.3 \pm 1.62$ & CPL & $268 \pm 48.5$ & -0.48 &\nodata& $0.77 \pm 0.13$ & $3.75 \pm 0.67$ & 1.75 & 128 & $-138.28 \pm 91.76$ & 32 & 10.9 & 3--10 & 17.6 &\nodata& 0 \\
		\hline
		Group D GRBs\\
		\hline
		150819440 & $0.96 \pm 0.09$ & CPL & $523.9 \pm 34.1$ & -1.14 &\nodata& $7.75 \pm 0.15$ & $148.44 \pm 3.77$ & 0.9 & 8 & $1.51 \pm 0.01$ & 0.7 & 2.5 & 0.682--10 & 73.8 &\nodata& 0 \\
		\enddata
		
		\tablenotetext{a}{The time resolution of the lightcurves used to calculate $\tau_{31}$.}
		\tablenotetext{b}{The probability that GRBs are classified as LGRBs in the $E_{\rm p}$--$T_{90}$ plane or the $HR$--$T_{90}$ plane.}
		\tablenotetext{c}{The redshift range that GRBs are classified as Type II GRBs in the $\epsilon$--$T_{\rm 90,z}$ plane or the $EH$--$T_{\rm 90,z}$ plane. The dotted lines represent that whatever the redshift is, it will not be classified as Type II GRBs.}
		\tablenotetext{d}{The probability that GRBs are classified as Type II GRBs in the $\epsilon$--$T_{\rm 90,z}$ plane or the $EH$--$T_{\rm 90,z}$ plane.}
		
	\end{deluxetable*}
\end{longrotatetable}

\end{document}